\documentclass[journal]{IEEEtran}
\usepackage{amsmath,amssymb,amsfonts,mathtools}
\usepackage{bm,graphicx,booktabs,multirow,array,tabularx}
\usepackage{cite,url,xcolor,tikz}
\usepackage{colortbl,placeins}
\usetikzlibrary{arrows.meta,positioning,calc,backgrounds}
\newcommand{\CAT}{WiFo-R}
\newcommand{\CC}{\mathbb C}
\newcommand{\RR}{\mathbb R}
\newcommand{\EE}{\mathbb E}
\newcommand{\LL}{\mathcal L}
\newcommand{\HH}{\mathcal H}
\newcommand{\YY}{\mathcal Y}
\newcommand{\MM}{\mathcal M}
\newcommand{\vv}[1]{\boldsymbol{#1}}

\definecolor{ink}{RGB}{32,70,83}
\definecolor{mint}{RGB}{232,242,240}
\definecolor{sand}{RGB}{250,240,227}
\definecolor{ice}{RGB}{230,240,251}
\newcolumntype{Y}{>{\raggedright\arraybackslash}X}
\begin{document}
\title{WiFo-R: Wireless Foundation Model with Recurrent Channel Representation}
\author{Xuanyu Liu,~\IEEEmembership{Graduate Student Member},~IEEE, Keyi Cao, and Xiang Cheng,~\IEEEmembership{Fellow,~IEEE}

\thanks{Xuanyu Liu, and Xiang Cheng are with the State Key Laboratory of Photonics and Communications, School of Electronics, Peking University, Beijing 100871, China (e-mail: xyliu25@stu.pku.edu.cn;  xiangcheng@pku.edu.cn).

Keyi Cao is with the School of Electronic Science and Engineering, Nanjing University, Nanjing 210023, China (e-mail: keyicao1@icloud.com)
}
}

\maketitle
\begin{abstract}
Wireless foundation models learn transferable channel state information (CSI) representations across heterogeneous wireless systems. However, conventional grid-aligned representations couple feature sequence length to CSI dimensions, restricting the allocation of representation capacity to different propagation conditions and task requirements while increasing inference overhead. This paper proposes WiFo-R, a wireless foundation model that learns progressive channel representations with selectable capacity. A channel-query cross-attention encoder aggregates spatial, temporal, and spectral CSI information, while parameter-shared recurrent updates refine existing channel features and introduce additional latent tokens. Multi-mask, multi-depth self-supervised pretraining equips intermediate states with strong reconstruction capabilities, followed by task-aligned post-training for channel estimation and CSI feedback. WiFo-R supports temporal and frequency-domain channel prediction, channel estimation, and variable-rate CSI feedback, and transfers to localization and LoS/NLoS classification. Experiments across in-distribution and out-of-distribution channel configurations demonstrate accurate reconstruction, progressive performance improvements, and strong transferability. A single feedback encoder--decoder supports six payload lengths at fixed quantization precision, while cross-attention and selectable recurrent depth improve the accuracy--complexity trade-off for large-scale CSI observations.
\end{abstract}


\section{Introduction}\label{sec:introduction}
\begin{figure}[!t]
\centering
\includegraphics[width=0.98\columnwidth]{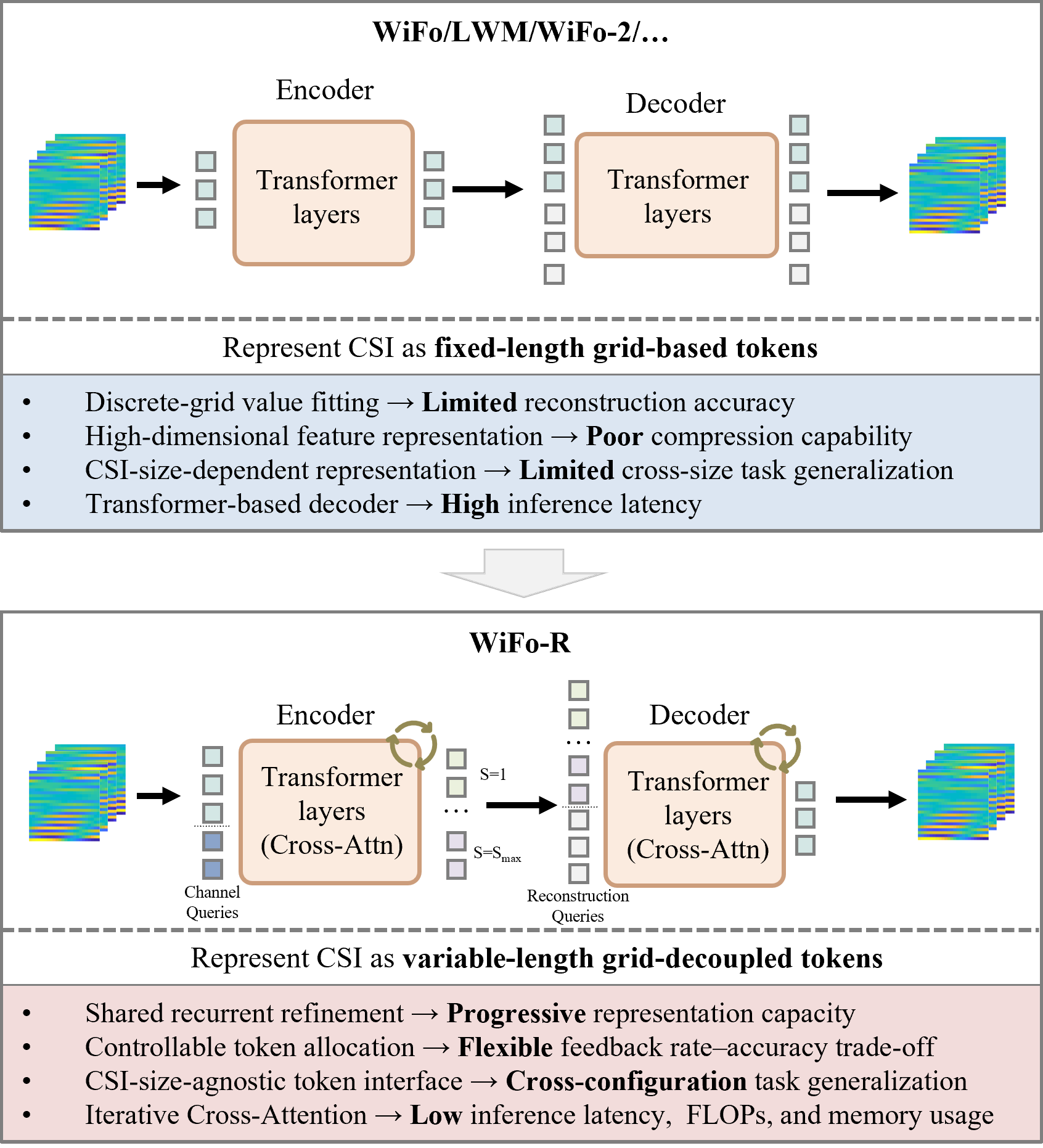}
\caption{Grid-aligned CSI tokenization versus recurrent channel representation in WiFo-R.}
\label{fig:overview}
\end{figure}

\IEEEPARstart{A}{ccurate} channel state information (CSI) is essential to a broad range of wireless communication and CSI-related functions, including channel prediction, channel estimation, CSI feedback, and CSI-based localization. Although these tasks differ in their observation patterns and desired outputs, they rely on common characteristics of radio propagation, such as spatial correlation, multipath delay structure, and temporal variation. Learning transferable channel representations is therefore an attractive alternative to developing separate models for individual tasks and operating conditions. Contemporary WiFo-series foundation models, including WiFo~\cite{liu2024wifo} and WiFo-2~\cite{liu2026wifo2}, have demonstrated the potential of large-scale CSI pre-training to support heterogeneous communication and channel-related applications. The effectiveness of such models nevertheless depends not only on the backbone architecture and pre-training objective but also on how CSI is converted into a sequence of features for subsequent processing.

A distinctive challenge in wireless representation learning arises because practical CSI observations exhibit two different sources of variability. The first arises from the \emph{acquisition configuration}: the numbers of antenna elements, subcarriers, and time samples may vary across devices, systems, and observation intervals, giving rise to CSI tensors of different dimensions. The second arises from the \emph{information to be represented}: channels with comparable observation dimensions can exhibit different multipath richness, delay and Doppler spreads, spatial correlations, and noise levels. Moreover, different tasks and target reconstruction accuracies may require different amounts of channel information. Importantly, these two forms of variability do not necessarily change together. Denser sampling of a channel over a similar physical aperture, bandwidth, or time interval can increase the number of measured CSI entries without a commensurate increase in its effective degrees of freedom; conversely, a fixed-size CSI tensor may describe channels of substantially different complexity.

This distinction is not fully reflected by conventional grid-based CSI tokenization. In typical patch-based Transformer architectures, each time--frequency--antenna patch is mapped to a feature token, so that the sequence length is determined by the CSI dimensions and the patching rule. The resulting token count therefore primarily reflects the \emph{sampling grid}, rather than the amount of information needed to characterize the current channel. When spatially, temporally, or spectrally correlated observations are represented by separate tokens, the backbone may repeatedly process redundant features. In particular, global self-attention over $N$ patch tokens incurs $O(N^2)$ attention-score interactions per layer. More fundamentally, a variable number of grid tokens complicates the construction of a compact, configuration-independent feature interface for downstream tasks. Simply increasing patch size or discarding a subset of patches reduces sequence length but may also eliminate important channel information.

As demonstrated by query-based latent modeling~\cite{jaegle2021perceiver,jaegle2022perceiverio}, a fixed-size latent bottleneck can decouple the output feature dimension from the input grid. However, imposing the same number of latent tokens on every observation introduces a different limitation: the representation capacity is fixed regardless of the propagation conditions or the requirements of the downstream task. An excessively large latent representation incurs unnecessary processing for comparatively simple channels, whereas an overly compact one may lose information that would be useful in more challenging scenarios. This motivates a channel representation that satisfies two complementary requirements: it should \emph{aggregate CSI globally from variable-size observations}, and it should \emph{provide different amounts of representational capacity within a single shared model}. The latter requirement calls for more than token pruning~\cite{rao2021dynamicvit,bolya2023tome}, since a shorter sequence should retain meaningful global channel information rather than correspond merely to a subset of the original sampling locations.

The proposed \CAT\ foundation model couples channel-query resampling with parameter-shared recurrent representation learning and multi-depth pretraining. Its encoder employs learnable channel queries to aggregate information from CSI patches through cross-attention, generating a compact representation whose length is not dictated by the input grid. A parameter-shared recurrent operator subsequently revisits the CSI features, updates the previously generated tokens, and introduces additional tokens. The representation thus evolves from a compact initial state toward increasingly expressive latent states. To maintain the usefulness of intermediate states rather than relying exclusively on the deepest one, the model adopts multi-budget masked channel-prediction pretraining across recurrent depths, followed by task-specific post-training for channel estimation and feedback. In this manner, the same model can provide channel features at several sequence lengths while preserving a common task interface.

The progressively refined representation is applicable to both CSI reconstruction and CSI-related tasks. For channel prediction and estimation, a shared reconstruction interface maps the learned latent features to the CSI positions of interest. For CSI compression feedback, representations produced at different recurrent depths can be converted into quantized codewords of different lengths and reconstructed by a shared feedback decoder. With multi-rate training, this provides a natural means of supporting different feedback rates using one model, while exploiting the same channel features learned for general-purpose representation. The resulting flexibility is distinct from designing a separate compression network for each rate. Beyond CSI reconstruction, compact channel representations can also be transferred to localization and line-of-sight/non-line-of-sight (LoS/NLoS) classification, demonstrating the transferability of the same representation to channel-related tasks.

The main contributions of this work are summarized as follows:
\begin{itemize}
\item \emph{A recurrent wireless foundation model for heterogeneous CSI:} We design a channel-query cross-attention encoder and a coordinate-conditioned decoder that learn global CSI representations independent of observation dimensions. Shared recurrent updates progressively refine existing channel knowledge and expand its representation capacity.
\item \emph{Multi-depth channel representation learning:} A self-supervised pretraining objective jointly learns spatial, temporal, and spectral dependencies while supervising multiple recurrent depths. Task-aligned post-training transfers the common backbone to channel estimation and channel feedback, and lightweight task heads extend its representations to channel-related tasks.
\item \emph{Unified multi-rate feedback and efficient cross-task inference:} The recurrent encoder--decoder supports six feedback payloads through feature-length adaptation without separate rate-specific models. Extensive ID and OOD experiments establish accuracy across four CSI reconstruction tasks, progressive downstream performance, and favorable accuracy--complexity scaling on large CSI observations.
\end{itemize}

\section{Related Work}\label{sec:related}
\subsection{Wireless Foundation Models and CSI Learning}
Deep learning has been applied to CSI estimation and prediction using convolutional, recurrent, and attention-based architectures~\cite{ye2018deepce,soltani2019channelnet,jiang2022transformer,luan2023channelformer}. LLM4CP~\cite{liu2024llm4cp} further adapts a pretrained language-model backbone to wireless channel prediction by introducing CSI-specific processing and adaptation modules. These methods demonstrate the effectiveness of learned channel features, but their models or fine-tuning procedures are primarily organized around particular prediction or estimation tasks. LLM4WM~\cite{liu2025llm4wm} extends pretrained-model adaptation to multiple channel-associated tasks through parameter-efficient multi-task learning.

Wireless-specific foundation models instead seek representations that transfer across propagation conditions, system configurations, and application tasks. LWM~\cite{alikhani2024lwm} studies large-scale wireless representation learning, while the contemporary WiFo family, including WiFo~\cite{liu2024wifo}, WiFo-2~\cite{liu2026wifo2}, WiFo-E~\cite{wen2026wifoe}, and WiFo-MiSAC~\cite{liu2026wifomisac}, extends pretrained channel features to prediction, precoding, and multimodal communication tasks.  MUSE-FM~\cite{zheng2026muse} further explores environment-aware representation learning for diverse wireless tasks. Broader frameworks for AI-native multimodal sensing--communication integration are developed in~\cite{cheng2024som,cheng2026fm4som,cheng2025synthsom}. These works establish the feasibility of a shared CSI backbone, but patch-based CSI tokenization still couples the number of intermediate grid tokens to the acquired tensor dimensions. WiFo-INR~\cite{liu2026wifoinr} addresses a related issue through compact, size-independent modulation tokens and coordinate-conditioned decoding. Our focus is complementary: instead of assigning a single fixed capacity to the channel representation, \CAT\ allows its number of latent tokens to increase progressively while maintaining the same pretrained backbone.

\subsection{Efficient Tokenization and Variable-Length Representations}
Reducing the number of tokens processed by a Transformer~\cite{vaswani2017attention} is a widely studied approach to improving computational efficiency. TokenLearner~\cite{ryoo2021tokenlearner}, DynamicViT~\cite{rao2021dynamicvit}, and Token Merging~\cite{bolya2023tome} respectively learn compact tokens, sparsify tokens, or merge redundant ones in visual recognition. The Perceiver architecture~\cite{jaegle2021perceiver} uses learnable latent queries and asymmetric cross-attention to process large input sequences through a compact latent array, and Perceiver IO~\cite{jaegle2022perceiverio} extends query-based processing to structured outputs. These methods motivate separating input resolution from the representation consumed by deeper network layers.

Although masked representation learning~\cite{he2022mae} and continuous neural representations~\cite{sitzmann2020siren} have broadened reconstruction modeling, a fixed-length latent array provides only one capacity. Adaptive computation~\cite{graves2016act,cai2020once} and ALIT~\cite{duggal2025alit} investigate recurrent token allocation in vision, adding new latent tokens while updating previously generated tokens at successive steps. These studies establish recurrent token allocation as an effective mechanism for constructing variable-length latent representations. In wireless modeling, ConsisFormer~\cite{wang2026consisformer} exploits local channel consistency to merge correlated CSI tokens and recover the feature sequence for downstream processing. \CAT\ addresses a different aspect of CSI tokenization: global cross-attention resampling provides a feature interface independent of the CSI grid, and recurrent refinement supplies multiple latent capacities from one model. The proposed learning and evaluation framework specifically considers time--frequency--antenna correlations, partial CSI observations, and transfer to reconstruction and channel-related tasks. At each depth, the recurrent operator refreshes the entire latent state while introducing new channel tokens.

\begin{figure*}[!t]
\centering
\includegraphics[width=0.985\textwidth]{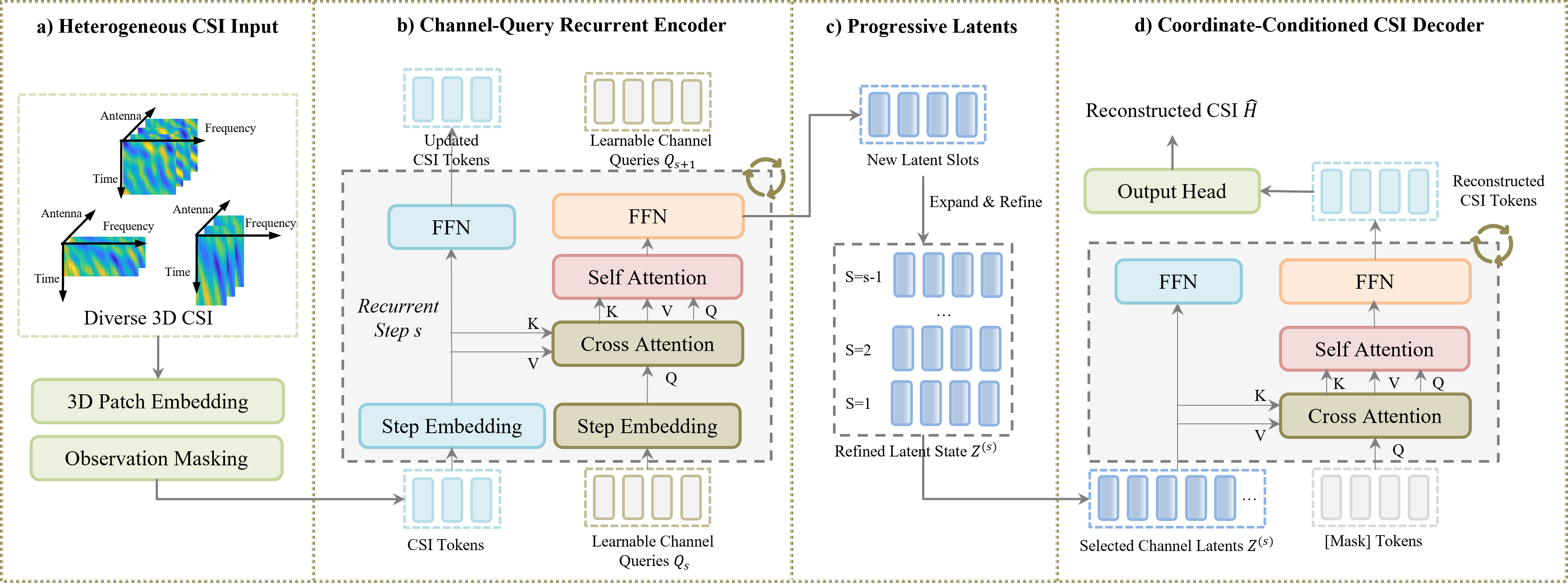}
\caption{Architecture of WiFo-R: (a) CSI embedding and masking, (b) recurrent channel-query encoding, (c) progressive latent refinement, and (d) coordinate-conditioned decoding.}
\label{fig:architecture}
\end{figure*}

\subsection{CSI Feedback and Channel-Related Tasks}
Neural CSI feedback compresses a high-dimensional downlink channel into a finite-rate representation and reconstructs the CSI at the receiver. CsiNet~\cite{wen2018csinet} introduced an autoencoder-based framework, while TransNet~\cite{cui2022transnet} exploited attention-based modeling for feedback reconstruction. Multiple-rate feedback networks~\cite{guo2020multirate,lu2020crnet} and architecture designs supporting configuration changes~\cite{lin2024scalable} further improve flexibility. More recently, WiFo-CF~\cite{liu2025wifocf} develops a wireless foundation model for heterogeneous CSI feedback with multi-user, multi-rate pretraining and shared/routed experts. It supports different feedback rates primarily by varying the quantization bit width applied to a compressed feature vector whose dimension is fixed by the encoder. In contrast, \CAT\ changes the \emph{number of compressed channel features} through recurrent-depth-controlled token allocation. This distinction is especially relevant when the quantization precision is held fixed: multiple bit budgets can then be realized with a shared set of recurrent encoder, compression, and reconstruction parameters instead of separately configured feature-length models. Beyond CSI reconstruction, the learned channel features also support localization and LoS/NLoS classification, for which wireless-specific readouts and task baselines such as WiT~\cite{salihu2022wit} are relevant.
\section{System Model and Problem Formulation}\label{sec:system}
\subsection{Heterogeneous CSI Observations}
Consider a MIMO--OFDM channel observed over $T$ time instants, $F$ frequency points, and $A=N_{\mathrm r}N_{\mathrm t}$ antenna pairs. The target complex-valued CSI is denoted by
\begin{equation}\label{eq:channel}
 \HH\in\CC^{T\times F\times A},\qquad A=N_{\mathrm r}N_{\mathrm t}.
\end{equation}
For physical interpretation, a local specular multipath approximation of the antenna-domain response at time $t$ and frequency $f$ is
\begin{equation}\label{eq:physical}
 \vv H(t,f)\simeq\sum_{\ell=1}^{P}\alpha_\ell
 e^{\mathrm j2\pi\nu_\ell t}e^{-\mathrm j2\pi f\tau_\ell}
 \vv a_{\mathrm r}(\vartheta_\ell^{\mathrm r})
 \vv a_{\mathrm t}^{\mathsf H}(\vartheta_\ell^{\mathrm t}),
\end{equation}
where $P$ is the number of resolved paths, $\alpha_\ell$, $\tau_\ell$, and $\nu_\ell$ are path gains, delays, and Doppler shifts, and $\vv a_{\mathrm r}$ and $\vv a_{\mathrm t}$ are array responses. The path-dependent delays and Doppler shifts induce structured, but potentially heterogeneous, correlations across the spatial, temporal, and frequency dimensions. Such structure explains why a finer observation grid does not necessarily require a proportionally larger learned channel representation.

The observation mask $\MM\in\{0,1\}^{T\times F\times A}$ specifies observed CSI positions and yields
\begin{equation}\label{eq:observation}
 \YY=\MM\odot(\HH+\vv E),
\end{equation}
where $\vv E$ represents measurement noise. The visible positions may be past time samples, available frequency bands, pilot locations, or the CSI estimate accessible to a feedback encoder.

\subsection{Unified Reconstruction and Channel-Related Tasks}
The observation pattern differs across tasks. Temporal prediction uses the available past CSI samples, frequency prediction uses an observed subband, and channel estimation relies on sparsely sampled pilot positions. CSI feedback further compresses the available channel observation into a rate-constrained codeword, while classification and localization map CSI to task-related labels. In each case, the reconstruction or inference target is denoted by $\vv y_\tau$, where $\tau$ identifies the corresponding task. An individual task is characterized by an observation operator $\mathcal O_\tau$, a target $\vv y_\tau$, and a reconstruction or prediction mapping $g_\tau$. The unified learning objective is
\begin{equation}\label{eq:systemobjective}
\min_{g_\tau}\;\EE_{\HH,\tau}\!\left[\mathcal L_\tau\!\left(g_\tau(\mathcal O_\tau(\HH)),\vv y_\tau\right)\right],
\end{equation}
where $\mathcal L_\tau$ is NMSE for channel reconstruction or a task-specific regression or classification loss. The acquisition grid, observation operator, and target may vary across configurations, motivating a transferable CSI representation.

\section{WiFo-R: Recurrent Channel Representation}\label{sec:method}
\subsection{Architecture Overview and CSI Embedding}
The shared model comprises a CSI patch embedder, a cross-attention token resampler, and a recurrent latent update; a reconstruction decoder is used for self-supervised learning, while lightweight readouts support subsequent tasks. A fixed-width patch embedder may be written as
\begin{equation}\label{eq:embedding}
 \vv x_i=W_{\mathrm e}
 [\Re\{\vv y_i\};\Im\{\vv y_i\};\vv m_i]
 +\vv\pi(\vv r_i,\vv c),
\end{equation}
where $\vv y_i$ is the available CSI in patch $i$, $\vv m_i$ indicates valid positions, and $\vv\pi$ encodes patch coordinates or other legitimate receiver-side metadata. Irregular patch boundaries are padded with a corresponding validity mask. By including the mask explicitly, unknown CSI values are not treated as physically observed zeros.

Let $\vv X=\{\vv x_i\}_{i=1}^{N}\in\RR^{N\times d}$ denote the input patch embeddings, with $N$ determined by the acquisition configuration and visibility mask. At recurrent depth $s$, WiFo-R generates $\vv Z^{(s)}\in\RR^{sK\times d}$, where $K$ channel queries are introduced per iteration and $s$ is an independently selectable representation budget. The encoder does not create one final feature token per input patch. Instead, $K$ learnable channel queries read the sequence $\vv X$ through cross-attention. The query bank is shared in width and processing parameters across recurrent depths, and the output is conditioned on global CSI observations rather than associated permanently with one location in the original grid. This is particularly relevant when a patch sequence contains correlated samples from the same physical propagation process.

\begin{figure}[!t]
\centering
\includegraphics[width=0.48\textwidth]{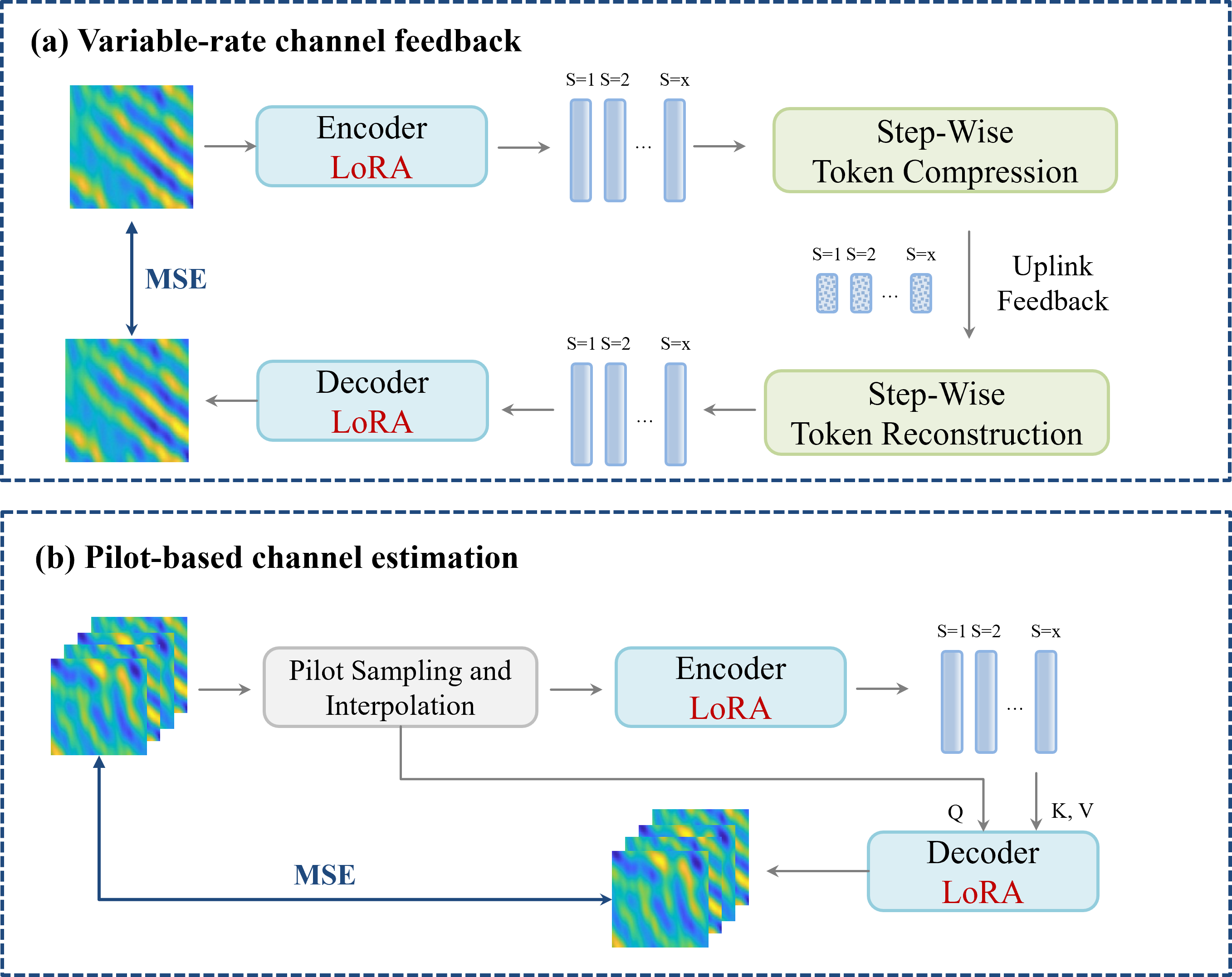}
\caption{Task-aligned post-training for (a) variable-rate CSI feedback and (b) pilot-aided channel estimation.}
\label{fig:posttrain}
\end{figure}

\subsection{Channel-Query Resampling and Recurrent Updates}
For multi-head cross-attention with a query sequence $\vv Q$ and key/value sequence $\vv V$, the $h$-th head computes
\begin{equation}\label{eq:ca}
 \mathrm{CA}_{h}(\vv Q,\vv V)=
 \mathrm{softmax}\!\left(
 \frac{(\vv QW_h^Q)(\vv VW_h^K)^\mathsf T}{\sqrt{d_h}}
 \right)\vv VW_h^V.
\end{equation}
The query outputs of all heads are projected to width $d$. The observed CSI supplies the keys and values, enabling a fixed set of latent queries to integrate evidence from a variable-size input. Unlike full input-grid self-attention, the query--key score matrix does not require all $N^2$ input-input interactions.

Let $\vv Q\in\RR^{K\times d}$ denote a learnable
channel-query bank shared across all recurrent steps.
At each step, learnable step embeddings are added
to the shared queries and the evolving CSI features,
with broadcasting along the token dimension.
Starting from an empty latent state $\vv Z^{(0)}$,
the recurrent update is formulated as
\begin{equation}\label{eq:rec}
\begin{aligned}
\vv U_s &=
[\vv Z^{(s-1)};\vv Q+\vv e_s^Q],
\quad
\vv V_s=\vv X^{(s-1)}+\vv e_s^X,\\
\vv Z^{(s)}
&=\mathcal R_\theta\bigl(
\vv U_s+
\mathrm{CA}_\theta(
\mathrm{LN}(\vv U_s),
\mathrm{LN}(\vv V_s))
\bigr).
\end{aligned}
\end{equation}
Here, $\vv e_s^Q$ and $\vv e_s^X$ denote
the step embeddings for the query and CSI
streams, respectively. The CSI features
$\vv X^{(s-1)}$ are progressively updated
by a shared feed-forward module, with
$\vv X^{(0)}=\vv X$.
Unlike the query sequence, the key/value
sequence contains only the evolving CSI
features, excluding previously generated
latent tokens. The shared recurrent
processor $\mathcal R_\theta$ jointly
refines existing and newly introduced
latent tokens through self-attention
and feed-forward operations.

\subsection{Progressive Masked CSI Representation Learning}
The foundation model is optimized through a CSI reconstruction interface that shares parameters across depths. For a masked target coordinate $\vv r_j^{\mathrm{out}}$, a decoder query reads $\vv Z^{(s)}$ and predicts the complex CSI sample as
\begin{equation}\label{eq:decoder}
 \hat h_j^{(s)}=W_{\mathrm o}\,\rho_\phi\!\left[
 \mathrm{CA}_\phi\!\left(e_{\mathrm{out}}(\vv r_j^{\mathrm{out}}),
 \vv Z^{(s)}\right)\right].
\end{equation}
The decoder query specifies the output sampling position; the output projection returns real and imaginary CSI components. The same decoder can process multiple latent lengths and requested CSI target positions. The query-based decoder is related to the structured output mechanism of Perceiver IO~\cite{jaegle2022perceiverio}, here specialized to the complex-valued CSI reconstruction targets.

\emph{Stage I: Multi-Depth Representation Pretraining.} Wireless reconstruction tasks exhibit substantial differences in both signal resolution and prediction targets. In particular, the coarser channel observations used in temporal and frequency prediction and the finer pilot grids required for CE exhibit markedly different space--time--frequency sampling characteristics. WiFo-R therefore separates learning shared CSI structure from task-specific adaptation. The initial stage performs self-supervised CSI prediction and reconstruction with three complementary masking patterns: random, temporal, and frequency masking. The masks vary across samples and recurrent depths, enabling the backbone to learn spatial, temporal, and spectral channel structure without depending on any fixed observation grid. For sample $i$, target set $\Omega_i$, and recurrent depth $s$, the reconstruction loss is
\begin{equation}\label{eq:recloss}
 \ell_i^{(s)}=\frac{\sum_{j\in\Omega_i}|\hat h_{i,j}^{(s)}-h_{i,j}|^2}
 {\sum_{j\in\Omega_i}|h_{i,j}|^2}.
\end{equation}
To make several operating depths usable within a shared model, a multi-budget channel-prediction objective is optimized
\begin{equation}\label{eq:progressive}
 \LL_{\mathrm{CP}}=\EE_{i,\tau\in\{\mathrm{CP\text{-}T},\mathrm{CP\text{-}F},\mathrm{RM}\}}\left[\sum_{s=1}^{S}\omega_s\ell_{i,\tau}^{(s)}\right], 
\end{equation}
where $\omega_s$ denotes the nonnegative loss weight at depth $s$, satisfying $\sum_{s=1}^{S}\omega_s=1$, and RM denotes random-masked CSI reconstruction. The expectation covers different channel observations and masking objectives. By placing reconstruction constraints on multiple recurrent states, the objective encourages early channel summaries to remain informative while later steps capture additional signal detail. As a result, the usefulness of an intermediate representation is learned explicitly rather than inherited from the final state.

\emph{Stage II: Task-Aligned Post-Training.} The pretrained recurrent channel representation is adapted to channel estimation (CE) and CSI feedback (CF) using the task-specific pathways in Fig.~\ref{fig:posttrain}. For CE, pilot sampling and interpolation provide the encoder input, and the decoder reconstructs CSI from the selected latent state. For CF, step-wise token compression and reconstruction surround a scalar-quantized feedback codeword. Low-rank adaptation (LoRA) aligns the pretrained encoder and decoder with each task, while the recurrent backbone and task-side modules share their weights across the supported operating depths. Both pathways are optimized by CSI reconstruction losses.

% \begin{figure}[!t]
% \centering
% \resizebox{0.98\columnwidth}{!}{\input{figures/posttraining_single_column.tex}}
% \caption{Stage-II post-training for (a) variable-rate CSI feedback and (b) channel estimation. Recurrent channel features are reused at selectable lengths; LoRA adapts the encoder and decoder, while task-specific interfaces implement quantized feedback or pilot-based reconstruction.}
% \label{fig:posttraining}
% \end{figure}

% Dataset configurations summarized from WiFo-2 Supplementary Tables S1--S3.
% Newly assigned WiFo-R identifiers are NOT WiFo-2 identifiers.
\begin{table*}[!t]
\centering
\caption{CSI Dataset Configurations for Training and Generalization Evaluation.}
\label{tab:settings}
\fontsize{6.85}{8.1}\selectfont
\setlength{\tabcolsep}{2.5pt}\renewcommand{\arraystretch}{1.10}
\begin{tabular*}{\textwidth}{@{\extracolsep{\fill}}llp{0.89in}ccccccp{1.33in}c@{}}
\toprule
\textbf{Usage} & \textbf{Source} & \textbf{Dataset} & \shortstack{$f_c$\\(GHz)} & \shortstack{$N_f$} & \shortstack{$\Delta f$\\(kHz)} & $T$ & \shortstack{$\Delta t$\\(ms)} & \textbf{Antenna} & \textbf{Scenario} & \shortstack{\textbf{Speed}\\(km/h)}\\
\midrule
\multirow{11}{*}{\textbf{Pre-training}} & \multirow{2}{*}{QuaDRiGa} & D1--D16 & 1.5--5.9 & 32--128 & 60--360 & 16--32 & 0.5--1 & $1\!\times\!4$--$4\!\times\!8$ & UMi/UMa/RMa/indoor; LoS/NLoS & 0--200\\
 & & D1CE--D16CE & 1.5--5.9 & 36--72 & 15--60 & 12--28 & 0.017--0.067 & $1\!\times\!4$--$4\!\times\!8$ & UMi/UMa/RMa/indoor; LoS/NLoS & 0--200\\
\cmidrule(l){2-11}
 & PKU-GBSM & VC, VF & 5.9 & 64--72 & 60--360 & 12--16 & 0.017--1 & $2\!\times\!4$--$4\!\times\!4$ & V2V & 36, 108\\
\cmidrule(l){2-11}
 & LuViRA & LC & 3.7 & 64 & 200 & 16 & 10 & 64 & Indoor & $\sim$0.5\\
\cmidrule(l){2-11}
 & DICHASUS & DC1--DC3 & 1.3--5.9 & 64 & 391 & 4 & -- & 64 & Indoor LoS/factory & --\\
\cmidrule(l){2-11}
 & Argos & AC1--AC5 & 2.4 & 52 & 313 & 4 & -- & 96 & Indoor/outdoor & --\\
\cmidrule(l){2-11}
 & \multirow{2}{*}{PKU-RM} & RMC1--RMC6 & 1.6--5.5 & 64 & 90--360 & 16 & 0.25--1 & 4 & Indoor/outdoor; LoS/NLoS & $\sim$2--5\\
 & & RMF1--RMF7 & 1.6--5.5 & 72 & 15--60 & 28 & 0.017--0.067 & 4 & Indoor/outdoor; LoS/NLoS & $\sim$2--5\\
\cmidrule(l){2-11}
 & \multirow{2}{*}{DeepMIMO} & DEC1 & 3.4 & 64 & 120 & 16 & 0.233 & 8 & Outdoor & 100--120\\
 & & DEF1--DEF2 & 3.4--28 & 36--72 & 30--120 & 12--28 & 0.008--0.022 & 8--$4\!\times\!8$ & Outdoor & 100--300\\
\cmidrule(l){2-11}
 & PKU-RT & RTC1--RTC2 & 5.9 & 64 & 180 & 16 & 2 & 8 & V2I & $\sim$10\\
\midrule
\multirow{7}{*}{\textbf{Generalization}} & \multirow{4}{*}{QuaDRiGa} & D17 & 3.5 & 64 & 60 & 16 & 0.5 & $2\!\times\!4$ & UMa NLoS & 30--100\\
 & & D18 & 6.7 & 64 & 60 & 16 & 1 & $4\!\times\!4$ & UMi LoS & 3--50\\
 & & D19 & 28 & 64 & 360 & 16 & 0.5 & $2\!\times\!8$ & UMa LoS & 30--100\\
 & & D17CE & 3.5 & 72 & 15 & 12 & 0.067 & $2\!\times\!4$ & UMa NLoS & 30--100\\
 & & D18CE & 6.7 & 36 & 15 & 12 & 0.067 & $4\!\times\!4$ & UMi LoS & 3--50\\
\cmidrule(l){2-11}
 & DeepMIMO & S1 & 28 & 32 & 480 & 16 & 0.029 & $4\!\times\!8$ & Outdoor & 100--300\\
\cmidrule(l){2-11}
 & Argos & A1 & 2.4 & 52 & 313 & 4 & -- & 96 & Outdoor & --\\
\cmidrule(l){2-11}
 & PKU-RM & R1 & 2.1 & 64 & 180 & 16 & 0.5 & 4 & Outdoor LoS & $\sim$5\\
\bottomrule
\end{tabular*}
\par\vspace{2pt}
\begin{minipage}{\textwidth}\fontsize{6.7}{8}\selectfont\raggedright
$N_f$ denotes the number of subcarriers; $T$ is the number of temporal samples. ``--'' indicates unavailable acquisition metadata.
\end{minipage}
\end{table*}

\subsection{Channel Reconstruction, Feedback, and Channel-Related Task Interfaces}\label{subsec:interfaces}
The shared coordinate-conditioned decoder reconstructs CSI at the requested temporal, frequency, and spatial sampling positions from the selected latent state. For temporal prediction, frequency-domain prediction, and channel estimation, the observation mask specifies the available CSI, while the decoder queries specify the target time, frequency, and antenna coordinates. The output dimensions can therefore change without requiring the latent sequence length to follow the observation grid.

\emph{Variable-rate CSI compression feedback} provides a further application of the progressive encoder--decoder structure. In an FDD system, the user equipment (UE) compresses the estimated downlink CSI, and the base station (BS) reconstructs it from the received codeword. For a latent state \(\vv Z^{(s)}\) and a prescribed payload of \(B\) bits, the feedback and reconstruction operations can be expressed as
\begin{subequations}\label{eq:feedback}\begin{align}
 \vv b^{(s,B)} &=\mathcal Q_B\!\left(\mathcal P_{\eta}(\vv Z^{(s)})\right)\in\{0,1\}^{B},\\
 \widehat{\HH}^{(s,B)}&=\mathcal D_{\varphi}\!\left(\vv b^{(s,B)},\mathcal C_{\rm out}\right),
\end{align}\end{subequations}
where \(\mathcal P_{\eta}\) is a feedback projection, \(\mathcal Q_B\) denotes the bit-constrained quantization and coding interface, and \(\mathcal C_{\rm out}\) specifies the reconstructed CSI coordinates. Multi-rate feedback training jointly optimizes the supported depths $s$ and bit budgets $B$, reusing the recurrent encoder and feedback decoder weights across all rates. At the fixed scalar precision of $b$ bits used in our experiments, a step-dependent projected vector of length $D_s$ yields $B=bD_s$ feedback bits. Increasing $s$ expands the channel features available to the step-wise compression and reconstruction modules, changing the feedback-vector length rather than the quantization precision. This provides variable-length feedback from a single trained model.

For CSI-based localization and LoS/NLoS classification, a task query aggregates the channel tokens into a fixed-width readout:
\begin{equation}\label{eq:readout}
 \vv a_\tau^{(s)}=\mathrm{Pool}\left[
 \mathrm{CA}_{\psi_\tau}(\vv Q_\tau,\vv Z^{(s)})\right],
\end{equation}
followed by a regression or classification head. Consequently, the same backbone accommodates reconstruction, compressed CSI transmission, and inference tasks with different target dimensions. The lightweight readouts operate directly on the compact latent interface, enabling channel-related task adaptation without full grid-length features.

\subsection{Computational Characteristics}\label{subsec:methodcost}
Let $N$ be the number of visible CSI patches, $K$ the newly introduced queries per iteration, $s$ the executed depth, and $d$ the hidden width. The cross-attention in~\eqref{eq:rec} uses $jK$ queries and $N+(j-1)K$ key/value tokens at iteration $j$. The cumulative number of attention query--key pairs is
\begin{align}\label{eq:cost}
 C_{\mathrm{CA}}(s)&=\sum_{j=1}^{s}jK[N+(j-1)K]\\
 &=\frac{NKs(s+1)}{2}
 +\frac{K^2s(s+1)(s-1)}{3}.\nonumber
\end{align}
For typical heterogeneous CSI observations, $K\ll N$ and the selected latent length $sK$ remains much smaller than the input patch sequence. Cross-attention therefore avoids the quadratic $N^2$ interactions of full grid self-attention, while parameter sharing avoids allocating a separate backbone for each depth. The compact latent array also reduces the amount of data consumed by downstream task heads. For example, a linear or MLP-based readout applied to channel tokens has an input-side cost proportional to $sK$ rather than the original $N$ grid tokens, yielding additional efficiency for CSI-based localization and classification. Increasing $s$ provides a controllable accuracy--complexity tradeoff without changing the pretrained weights.

\section{Experimental Evaluation}\label{sec:experiments}
This section evaluates WiFo-R as a unified wireless foundation model across four CSI reconstruction tasks: temporal channel prediction (CP-T), frequency-domain channel prediction (CP-F), channel estimation (CE), and CSI feedback (CF). We examine ID reconstruction accuracy, OOD generalization, progressive representation quality, and variable-rate feedback using shared parameters. Then, we assess the transferability of learned channel information through localization and LoS/NLoS classification, followed by inference-complexity measurements and ablation studies.

\subsection{Experimental Setup}\label{subsec:setup}
\emph{Datasets and evaluation protocol:} We use the heterogeneous LH-CSI collection adopted by WiFo-2~\cite{liu2026wifo2}. Its eight sources include statistical channel models, ray tracing, and real-world measurements, covering the frequency bands, array configurations, CSI sampling resolutions, and propagation environments summarized in Table~\ref{tab:settings}. QuaDRiGa D1--D16 provide channel-prediction pretraining data, with fine-grained D1CE--D16CE used for CE adaptation; other training sources broaden the channel statistics encountered during pretraining. The designated D17--D18, S1, A1, and R1 configurations are held out for task-dependent zero-shot evaluation, while feedback additionally uses the D17--D19 held-out split. No target-configuration training is performed when reporting these generalization results.

\emph{Implementation details:} 
The model adopts a patch size of $4\times4\times4$ along the temporal, spatial, and frequency dimensions, with a hidden dimension of $d=768$ and eight cross-attention heads. The recurrent encoder comprises two Transformer blocks with parameters shared across six recurrent steps. Each step introduces $K=16$ channel queries, yielding representations of 16--96 latent tokens for $s=1,\ldots,6$. The reconstruction decoder consists of a single Transformer block recurrently applied six times with shared parameters.

\emph{Training details:} 
Stage-I pretraining is conducted for 500 epochs, followed by separate Stage-II post-training for CE and CF, each for 500 epochs. All stages use a batch size of 32 and the Adam optimizer with $\beta_1=0.9$ and $\beta_2=0.999$. The learning rate follows a cosine annealing schedule with a 10-epoch warm-up, ranging from $1\times10^{-6}$ to $1\times10^{-5}$. Pretraining combines random-masked reconstruction, temporal CSI prediction (CP-T), and frequency-domain CSI prediction (CP-F), with SNR uniformly sampled from 10 to 25 dB and masking ratios from $10\%$ to $50\%$. The multi-depth reconstruction loss uses uniform weights $\omega_s=1/6$ for $s=1,\ldots,6$. CE post-training uses pilot strides sampled between $(4,1,6)$ and $(8,1,24)$, whereas CF post-training jointly covers the evaluated feedback lengths with five-bit scalar quantization. Unless stated otherwise, reconstruction testing uses 20-dB SNR, a $25\%$ prediction mask, and a CE pilot stride of $(4,1,12)$.

\emph{Baselines and metrics:} We compare conventional extrapolation/interpolation, task-specific neural reconstruction models, and pretrained wireless models, as identified in the tables. WiFo-2 Base-Dense is the principal pretrained reference for CP-T, CP-F, and CE; CsiNet, TransNet, and WiFo-CF provide complementary feedback comparisons. For CSI reconstruction, sample-level normalized error is
\begin{align}\label{eq:nmse}
 e_i&=\frac{\sum_{j\in\Omega_i}|h_{i,j}-\hat h_{i,j}|^2}{\sum_{j\in\Omega_i}|h_{i,j}|^2},\\
 \mathrm{NMSE}_{\mathrm{dB}}&=10\log_{10}\!\left(\frac{1}{n}\sum_{i=1}^{n}e_i\right).\nonumber
\end{align}
The summary tables report averages of dataset-level NMSE values in dB and average per-dataset method ranks. Localization is evaluated by position RMSE, LoS/NLoS classification by macro-F1, and computational demand by FLOPs and measured batch-one inference latency.

\subsection{Progressive CSI Reconstruction and OOD Generalization}\label{subsec:reconstruction}
We first evaluate the shared foundation model on CP-T, CP-F, and CE. CP-T predicts future CSI from past observations, CP-F reconstructs unobserved subcarriers, and CE recovers channel coefficients from sparse pilots. Tables~\ref{tab:reconstruction-id} and~\ref{tab:reconstruction-ood} report the ID and OOD results, respectively. We compare recurrent depths $s\in\{1,3,5\}$, corresponding to 16, 48, and 80 latent tokens, without changing the pretrained backbone parameters.

% Requires booktabs, array, xcolor, tikz, and \usetikzlibrary{backgrounds}. Compile at least twice. Source values in dB; average before rounding (ROUND_HALF_UP).
\begin{table*}[t]
\centering
\caption{In-distribution CSI reconstruction performance (NMSE in dB). Best/second-best values are red bold/blue underlined.}
\label{tab:reconstruction-id}
\fontsize{6.5}{7.8}\selectfont\setlength{\tabcolsep}{1.2pt}\renewcommand{\arraystretch}{1.12}
\begin{tikzpicture}[remember picture]
\node[anchor=west,inner sep=0pt,outer sep=0pt] (id-grid) {\begin{tabular*}{\textwidth}{@{\extracolsep{\fill}}l*{16}{r}w{c}{7.5mm}w{c}{7mm}@{}}
\toprule
\textbf{Method} & D1 & D2 & D3 & D4 & D5 & D6 & D7 & D8 & D9 & D10 & D11 & D12 & D13 & D14 & D15 & D16 & \makebox[7.5mm][c]{\tikz[remember picture,overlay]\coordinate (id-avg-0);\makebox[0pt][c]{\textbf{Avg.}}} & \makebox[7mm][c]{\tikz[remember picture,overlay]\coordinate (id-rank-0);\makebox[0pt][c]{\shortstack{\textbf{Avg.}\\\textbf{rank}}}} \\
\midrule
\multicolumn{19}{@{}l}{\emph{Temporal channel prediction}}\\
Extrapolation & 3.88 & 2.86 & -20.03 & 1.86 & 3.48 & 3.71 & 4.58 & -14.39 & 3.72 & 4.24 & 4.20 & -5.36 & 3.68 & 3.38 & -3.18 & 3.61 & 0.02 & 6.94 \\
Transformer & -9.31 & -5.28 & -19.76 & -5.82 & -7.46 & -6.81 & -2.45 & -15.82 & -6.85 & -5.09 & -2.47 & -14.62 & -2.78 & -5.70 & -15.11 & -1.75 & -7.94 & 6.06 \\
LLM4CP & -12.55 & -7.70 & -23.58 & -13.39 & -8.42 & -9.39 & -9.81 & -20.76 & -9.23 & -7.37 & -7.75 & -18.16 & -5.89 & -8.08 & -15.20 & -6.52 & -11.49 & 4.94 \\
WiFo-2-Base-Dense & -13.73 & -10.07 & -23.61 & \textcolor{blue}{\underline{-17.92}} & -10.78 & -12.98 & \textcolor{blue}{\underline{-14.84}} & -23.20 & -10.63 & -10.76 & -10.64 & -22.38 & -8.69 & -10.61 & -21.58 & -9.21 & -14.48 & 3.38 \\
WiFo-R ($s=1$) & -14.99 & -10.01 & -23.67 & -12.95 & -12.97 & -12.63 & -10.53 & -23.58 & -11.89 & -10.80 & -9.00 & -21.13 & -7.26 & -10.70 & -22.44 & -7.78 & -13.90 & 3.56 \\
WiFo-R ($s=3$) & \textcolor{blue}{\underline{-16.53}} & \textcolor{blue}{\underline{-11.65}} & \textcolor{blue}{\underline{-26.13}} & -17.87 & \textcolor{blue}{\underline{-14.89}} & \textcolor{blue}{\underline{-14.47}} & -14.59 & \textcolor{blue}{\underline{-26.10}} & \textcolor{blue}{\underline{-13.40}} & \textcolor{blue}{\underline{-12.47}} & \textcolor{blue}{\underline{-11.17}} & \textcolor{blue}{\underline{-24.29}} & \textcolor{blue}{\underline{-9.34}} & \textcolor{blue}{\underline{-12.33}} & \textcolor{blue}{\underline{-24.35}} & \textcolor{blue}{\underline{-10.09}} & \textcolor{blue}{\underline{-16.23}} & \textcolor{blue}{\underline{2.13}} \\
WiFo-R ($s=5$) & \textcolor{red}{\textbf{-16.76}} & \textcolor{red}{\textbf{-11.84}} & \textcolor{red}{\textbf{-26.66}} & \textcolor{red}{\textbf{-18.87}} & \textcolor{red}{\textbf{-15.04}} & \textcolor{red}{\textbf{-14.72}} & \textcolor{red}{\textbf{-15.42}} & \textcolor{red}{\textbf{-26.80}} & \textcolor{red}{\textbf{-13.57}} & \textcolor{red}{\textbf{-12.75}} & \textcolor{red}{\textbf{-11.71}} & \textcolor{red}{\textbf{-24.76}} & \textcolor{red}{\textbf{-9.84}} & \textcolor{red}{\textbf{-12.54}} & \textcolor{red}{\textbf{-24.68}} & \textcolor{red}{\textbf{-10.49}} & \textcolor{red}{\textbf{-16.65}} & \textcolor{red}{\textbf{1.00}} \\
\midrule
\multicolumn{19}{@{}l}{\emph{Frequency-domain channel prediction}}\\
Extrapolation & 2.50 & 0.82 & -6.85 & -0.91 & -3.46 & 0.21 & -0.22 & 0.07 & 2.03 & -0.35 & 2.44 & -6.88 & 2.27 & 1.87 & -6.86 & 1.95 & -0.71 & 7.00 \\
Transformer & -4.97 & -1.72 & -16.21 & -4.09 & -7.88 & -4.92 & -6.98 & -10.91 & -4.43 & -3.50 & -4.52 & -12.34 & -3.52 & -2.61 & -12.30 & -6.79 & -6.73 & 5.94 \\
LLM4CP & -8.31 & -8.35 & -19.58 & -7.96 & -11.93 & -9.87 & -6.48 & -15.31 & -7.79 & -6.88 & -5.08 & -16.40 & -4.18 & -7.15 & -19.33 & -7.13 & -10.11 & 5.06 \\
WiFo-2-Base-Dense & -10.30 & \textcolor{red}{\textbf{-13.48}} & -22.12 & \textcolor{red}{\textbf{-15.13}} & -16.97 & \textcolor{red}{\textbf{-14.86}} & \textcolor{red}{\textbf{-13.69}} & -18.46 & -10.26 & \textcolor{red}{\textbf{-13.54}} & \textcolor{red}{\textbf{-10.81}} & -21.88 & \textcolor{red}{\textbf{-10.14}} & -10.77 & -22.43 & \textcolor{red}{\textbf{-12.54}} & \textcolor{blue}{\underline{-14.84}} & \textcolor{blue}{\underline{2.00}} \\
WiFo-R ($s=1$) & -9.74 & -11.27 & -21.75 & -11.52 & -16.07 & -11.76 & -9.13 & -17.82 & -9.52 & -10.38 & -7.68 & -20.31 & -6.54 & -9.33 & -22.18 & -7.33 & -12.65 & 4.00 \\
WiFo-R ($s=3$) & \textcolor{blue}{\underline{-11.52}} & -13.21 & \textcolor{red}{\textbf{-23.67}} & -13.56 & \textcolor{blue}{\underline{-18.10}} & -13.51 & -11.47 & \textcolor{blue}{\underline{-19.62}} & \textcolor{blue}{\underline{-11.07}} & -11.87 & -9.58 & \textcolor{blue}{\underline{-23.34}} & -8.16 & \textcolor{blue}{\underline{-10.92}} & \textcolor{blue}{\underline{-24.38}} & -11.21 & -14.70 & 2.44 \\
WiFo-R ($s=5$) & \textcolor{red}{\textbf{-11.63}} & \textcolor{blue}{\underline{-13.45}} & \textcolor{blue}{\underline{-23.66}} & \textcolor{blue}{\underline{-13.78}} & \textcolor{red}{\textbf{-18.29}} & \textcolor{blue}{\underline{-13.73}} & \textcolor{blue}{\underline{-11.71}} & \textcolor{red}{\textbf{-19.78}} & \textcolor{red}{\textbf{-11.16}} & \textcolor{blue}{\underline{-12.01}} & \textcolor{blue}{\underline{-9.80}} & \textcolor{red}{\textbf{-23.54}} & \textcolor{blue}{\underline{-8.32}} & \textcolor{red}{\textbf{-11.07}} & \textcolor{red}{\textbf{-24.55}} & \textcolor{blue}{\underline{-11.55}} & \textcolor{red}{\textbf{-14.88}} & \textcolor{red}{\textbf{1.56}} \\
\midrule
\multicolumn{19}{@{}l}{\emph{Channel estimation}}\\
Interpolation & -15.30 & -15.12 & -17.78 & -8.40 & -13.82 & -16.98 & -6.25 & -17.13 & -11.76 & -9.56 & -7.18 & -17.32 & -5.91 & -12.33 & -17.38 & -7.77 & -12.50 & 7.00 \\
Channelformer & -19.41 & -18.27 & -24.25 & -10.42 & -16.58 & -20.38 & -8.40 & -20.40 & -15.01 & -12.00 & -10.09 & -22.02 & -11.72 & -15.44 & -20.54 & -10.63 & -15.97 & 5.25 \\
LLM4CE & -18.64 & -16.57 & -20.63 & -10.22 & -15.95 & -19.23 & -8.22 & -21.01 & -14.26 & -11.54 & -10.60 & -20.42 & -11.62 & -14.98 & -21.64 & -10.84 & -15.40 & 5.75 \\
WiFo-2-Base-Dense & -25.70 & -23.49 & -30.24 & -13.93 & -23.04 & -27.98 & -10.52 & \textcolor{blue}{\underline{-26.17}} & -19.33 & -13.89 & -14.54 & -27.75 & -16.27 & -20.38 & -26.76 & -14.98 & -20.94 & 3.81 \\
WiFo-R ($s=1$) & -25.76 & -24.74 & -30.77 & -15.84 & -23.89 & -27.92 & -12.97 & -25.71 & -20.70 & -16.53 & -16.24 & -27.98 & -18.19 & -20.83 & -26.90 & -15.81 & -21.92 & 3.13 \\
WiFo-R ($s=3$) & \textcolor{blue}{\underline{-26.49}} & \textcolor{blue}{\underline{-25.81}} & \textcolor{blue}{\underline{-31.53}} & \textcolor{blue}{\underline{-16.82}} & \textcolor{blue}{\underline{-24.79}} & \textcolor{blue}{\underline{-28.87}} & \textcolor{blue}{\underline{-13.82}} & -26.15 & \textcolor{blue}{\underline{-21.51}} & \textcolor{blue}{\underline{-17.27}} & \textcolor{blue}{\underline{-16.95}} & \textcolor{blue}{\underline{-28.66}} & \textcolor{blue}{\underline{-19.04}} & \textcolor{blue}{\underline{-21.63}} & \textcolor{blue}{\underline{-27.26}} & \textcolor{blue}{\underline{-16.51}} & \textcolor{blue}{\underline{-22.70}} & \textcolor{blue}{\underline{2.06}} \\
WiFo-R ($s=5$) & \textcolor{red}{\textbf{-26.60}} & \textcolor{red}{\textbf{-25.94}} & \textcolor{red}{\textbf{-31.61}} & \textcolor{red}{\textbf{-16.94}} & \textcolor{red}{\textbf{-24.89}} & \textcolor{red}{\textbf{-29.00}} & \textcolor{red}{\textbf{-13.86}} & \textcolor{red}{\textbf{-26.22}} & \textcolor{red}{\textbf{-21.58}} & \textcolor{red}{\textbf{-17.33}} & \textcolor{red}{\textbf{-17.00}} & \textcolor{red}{\textbf{-28.76}} & \textcolor{red}{\textbf{-19.11}} & \textcolor{red}{\textbf{-21.75}} & \textcolor{red}{\textbf{-27.31}} & \textcolor{red}{\textbf{-16.62}} & \textcolor{red}{\textbf{-22.78}} & \textcolor{red}{\textbf{1.00}} \\
\bottomrule
\end{tabular*}};
\begin{scope}[overlay,on background layer]
\fill[black!10] ([xshift=\dimexpr-3.75mm-\tabcolsep\relax]id-avg-0 |- id-grid.north) rectangle ([xshift=\dimexpr3.75mm+\tabcolsep\relax]id-avg-0 |- id-grid.south);
\fill[red!8] ([xshift=\dimexpr-3.5mm-\tabcolsep\relax]id-rank-0 |- id-grid.north) rectangle ([xshift=\dimexpr3.5mm+\tabcolsep\relax]id-rank-0 |- id-grid.south);
\end{scope}
\end{tikzpicture}
 
\end{table*}

% Requires booktabs, array, xcolor, tikz, and \usetikzlibrary{backgrounds}. Compile at least twice. Dataset groups follow the requested task-specific split.
% CE Transformer in the CSV is displayed as Channelformer.
\begin{table*}[t]
\centering
\caption{Out-of-distribution CSI reconstruction performance (NMSE in dB). Best/second-best values are red bold/blue underlined.}
\label{tab:reconstruction-ood}
\fontsize{6.5}{7.8}\selectfont\setlength{\tabcolsep}{0.6pt}\renewcommand{\arraystretch}{1.12}
\begin{tikzpicture}[remember picture]
\node[anchor=west,inner sep=0pt,outer sep=0pt] (ood-grid) {\begin{tabular*}{\textwidth}{@{\extracolsep{\fill}}l*{4}{r}w{c}{7.5mm}w{c}{7mm}@{\hspace{4pt}}l*{5}{r}w{c}{7.5mm}w{c}{7mm}@{\hspace{4pt}}l*{2}{r}w{c}{7.5mm}w{c}{7mm}@{}}
\toprule
\multicolumn{7}{c}{\textbf{Temporal channel prediction}} & \multicolumn{8}{c}{\textbf{Frequency-domain channel prediction}} & \multicolumn{5}{c}{\textbf{Channel estimation}} \\
\cmidrule(lr){1-7}\cmidrule(lr){8-15}\cmidrule(l){16-20}
\textbf{Method} & D17 & D18 & \shortstack{R1} & \shortstack{S1} & \makebox[7.5mm][c]{\tikz[remember picture,overlay]\coordinate (ood-avg-0);\makebox[0pt][c]{\textbf{Avg.}}} & \makebox[7mm][c]{\tikz[remember picture,overlay]\coordinate (ood-rank-0);\makebox[0pt][c]{\shortstack{\textbf{Avg.}\\\textbf{rank}}}} & \textbf{Method} & D17 & D18 & \shortstack{A1} & \shortstack{R1} & \shortstack{S1} & \makebox[7.5mm][c]{\tikz[remember picture,overlay]\coordinate (ood-avg-1);\makebox[0pt][c]{\textbf{Avg.}}} & \makebox[7mm][c]{\tikz[remember picture,overlay]\coordinate (ood-rank-1);\makebox[0pt][c]{\shortstack{\textbf{Avg.}\\\textbf{rank}}}} & \textbf{Method} & D17CE & D18CE & \makebox[7.5mm][c]{\tikz[remember picture,overlay]\coordinate (ood-avg-2);\makebox[0pt][c]{\textbf{Avg.}}} & \makebox[7mm][c]{\tikz[remember picture,overlay]\coordinate (ood-rank-2);\makebox[0pt][c]{\shortstack{\textbf{Avg.}\\\textbf{rank}}}} \\
\midrule
Extrapolation & 4.91 & 3.98 & -18.50 & -15.26 & -6.22 & 6.75 & Extrapolation & 2.56 & 0.37 & 4.90 & 0.05 & -11.94 & -0.81 & 7.00 & Interpolation & -8.04 & -15.32 & -11.68 & 6.00 \\
Transformer & -4.05 & -6.85 & -17.42 & -16.40 & -11.18 & 6.25 & Transformer & -1.27 & -8.72 & -1.17 & -11.93 & -17.23 & -8.07 & 6.00 & Channelformer & -6.64 & -6.08 & -6.36 & 7.00 \\
LLM4CP & -8.63 & -10.78 & \textcolor{red}{\textbf{-21.17}} & -18.24 & -14.71 & 4.00 & LLM4CP & -5.26 & -9.11 & \textcolor{red}{\textbf{-14.03}} & -16.10 & -18.74 & -12.65 & 4.00 & LLM4CE & -10.90 & -17.78 & -14.34 & 5.00 \\
WiFo-2-Base-Dense & -12.27 & -13.86 & \textcolor{blue}{\underline{-20.57}} & -20.38 & -16.77 & 3.25 & WiFo-2-Base-Dense & \textcolor{red}{\textbf{-10.84}} & -14.18 & -12.91 & -14.21 & -22.05 & -14.84 & 3.40 & WiFo-2-Base-Dense & -14.89 & -24.26 & -19.58 & 3.50 \\
\raisebox{-0.45\height}{\shortstack[l]{WiFo-R ($s=1$)}} & -11.79 & -13.91 & -19.11 & -23.27 & -17.02 & 3.25 & \raisebox{-0.45\height}{\shortstack[l]{WiFo-R ($s=1$)}} & -8.17 & -12.68 & -9.88 & -16.65 & -24.09 & -14.29 & 3.80 & \raisebox{-0.45\height}{\shortstack[l]{WiFo-R ($s=1$)}} & -16.11 & -24.01 & -20.06 & 3.50 \\
\raisebox{-0.45\height}{\shortstack[l]{WiFo-R ($s=3$)}} & \textcolor{blue}{\underline{-13.39}} & \textcolor{blue}{\underline{-15.91}} & -19.09 & \textcolor{red}{\textbf{-23.74}} & \textcolor{blue}{\underline{-18.03}} & \textcolor{red}{\textbf{2.25}} & \raisebox{-0.45\height}{\shortstack[l]{WiFo-R ($s=3$)}} & -9.86 & \textcolor{blue}{\underline{-14.73}} & -13.03 & \textcolor{red}{\textbf{-16.93}} & \textcolor{red}{\textbf{-25.54}} & \textcolor{blue}{\underline{-16.02}} & \textcolor{blue}{\underline{2.00}} & \raisebox{-0.45\height}{\shortstack[l]{WiFo-R ($s=3$)}} & \textcolor{blue}{\underline{-17.02}} & \textcolor{blue}{\underline{-24.72}} & \textcolor{blue}{\underline{-20.87}} & \textcolor{blue}{\underline{2.00}} \\
\raisebox{-0.45\height}{\shortstack[l]{WiFo-R ($s=5$)}} & \textcolor{red}{\textbf{-13.66}} & \textcolor{red}{\textbf{-16.31}} & -19.00 & \textcolor{blue}{\underline{-23.45}} & \textcolor{red}{\textbf{-18.10}} & \textcolor{red}{\textbf{2.25}} & \raisebox{-0.45\height}{\shortstack[l]{WiFo-R ($s=5$)}} & \textcolor{blue}{\underline{-10.00}} & \textcolor{red}{\textbf{-14.93}} & \textcolor{blue}{\underline{-13.60}} & \textcolor{blue}{\underline{-16.88}} & \textcolor{blue}{\underline{-25.16}} & \textcolor{red}{\textbf{-16.11}} & \textcolor{red}{\textbf{1.80}} & \raisebox{-0.45\height}{\shortstack[l]{WiFo-R ($s=5$)}} & \textcolor{red}{\textbf{-17.13}} & \textcolor{red}{\textbf{-24.81}} & \textcolor{red}{\textbf{-20.97}} & \textcolor{red}{\textbf{1.00}} \\
\bottomrule
\end{tabular*}};
\begin{scope}[overlay,on background layer]
\coordinate (ood-band-top) at ([yshift=\dimexpr\ht\strutbox+\baselineskip\relax]ood-avg-0);
\fill[black!10] ([xshift=\dimexpr-3.75mm-\tabcolsep\relax]ood-avg-0 |- ood-band-top) rectangle ([xshift=\dimexpr3.75mm+\tabcolsep\relax]ood-avg-0 |- ood-grid.south);
\fill[red!8] ([xshift=\dimexpr-3.5mm-\tabcolsep\relax]ood-rank-0 |- ood-band-top) rectangle ([xshift=\dimexpr3.5mm+\tabcolsep\relax]ood-rank-0 |- ood-grid.south);
\fill[black!10] ([xshift=\dimexpr-3.75mm-\tabcolsep\relax]ood-avg-1 |- ood-band-top) rectangle ([xshift=\dimexpr3.75mm+\tabcolsep\relax]ood-avg-1 |- ood-grid.south);
\fill[red!8] ([xshift=\dimexpr-3.5mm-\tabcolsep\relax]ood-rank-1 |- ood-band-top) rectangle ([xshift=\dimexpr3.5mm+\tabcolsep\relax]ood-rank-1 |- ood-grid.south);
\fill[black!10] ([xshift=\dimexpr-3.75mm-\tabcolsep\relax]ood-avg-2 |- ood-band-top) rectangle ([xshift=\dimexpr3.75mm+\tabcolsep\relax]ood-avg-2 |- ood-grid.south);
\fill[red!8] ([xshift=\dimexpr-3.5mm-\tabcolsep\relax]ood-rank-2 |- ood-band-top) rectangle ([xshift=\dimexpr3.5mm+\tabcolsep\relax]ood-rank-2 |- ood-grid.south);
\end{scope}
\end{tikzpicture}
 
\end{table*}

At $s=5$, WiFo-R achieves ID NMSEs of $-16.65$ dB for CP-T, $-14.88$ dB for CP-F, and $-22.78$ dB for CE. Compared with WiFo-2 Base-Dense, the respective gains are $2.17$, $0.04$, and $1.84$ dB. On OOD configurations, WiFo-R attains $-18.10$, $-16.11$, and $-20.97$ dB, improving over WiFo-2 by $1.33$, $1.27$, and $1.39$ dB. These results establish accurate reconstruction across heterogeneous observation operators and demonstrate generalization to unseen wireless environments.

Reconstruction quality also improves as the recurrent representation expands. From $s=1$ to $s=5$, the ID NMSE decreases by $2.75$ dB for CP-T, $2.23$ dB for CP-F, and $0.86$ dB for CE; the OOD averages improve by $1.08$, $1.82$, and $0.91$ dB, respectively. The larger early-depth gains in CP-T and CE, together with continued improvements in CP-F, are consistent with progressive extraction of dominant channel structure followed by finer spectral details. The same learned parameters thus provide multiple accuracy--capacity operating points for different reconstruction demands.

\begin{figure}[t]
\centering
\includegraphics[width=\columnwidth]{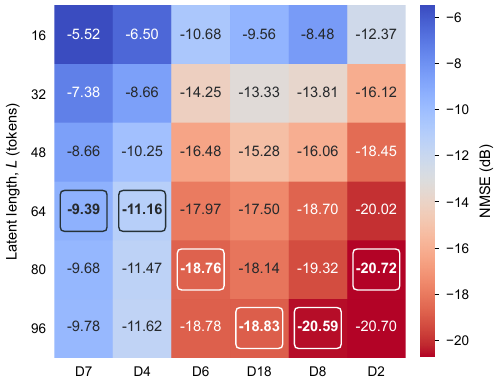}
\caption{CSI reconstruction NMSE across latent lengths and channel configurations. Outlined cells indicate the shortest length within 0.5 dB of the best NMSE for each configuration.}
\label{fig:capacity}
\end{figure}

To further investigate the dependence of representation capacity on channel characteristics, we consider CSI feedback, where the full CSI is reconstructed from compact latent representations. The reconstruction NMSE thus provides a direct measure of how well channel information is preserved under different representation budgets. Fig.~\ref{fig:capacity} evaluates six datasets with an identical input size of 2048 real-valued CSI elements. We evaluate latent lengths from 16 to 96 tokens and identify the shortest length within $0.5$ dB of each dataset's best measured NMSE. D4 and D7 reach this accuracy range with 64 tokens, D2 and D6 with 80, and D8 and D18 with 96. Different channel configurations therefore require different representation capacities despite having the same input dimensions, demonstrating the benefit of controlling channel representation length independently of the CSI grid.

\subsection{Variable-Rate CSI Feedback}\label{subsec:feedback}
We next evaluate CSI feedback, the fourth reconstruction task, to assess reconstruction accuracy and rate flexibility within one shared model. The feedback encoder converts recurrent channel representations into quantized features, and the decoder reconstructs CSI from these features. Six nominal payloads from 160 to 960 bits are jointly supported by varying the feature-vector length at a fixed precision of five bits per scalar. Table~\ref{tab:feedback} compares WiFo-R with CsiNet, TransNet, and WiFo-CF on the ID configurations D1--D16 and the OOD configurations D17--D19.

% Feedback values are preserved from the previous manuscript.
% values and are excluded from the best/second-best formatting.
% Visual convention matches Tables II--III: grey average columns,
% red bold for best and blue underlining for second best (measured methods).
\begin{table*}[t]
\centering
\caption{Variable-rate CSI feedback performance (NMSE in dB). Best/second-best values are red bold/blue underlined.}
\label{tab:feedback}
\fontsize{6.5}{7.8}\selectfont\setlength{\tabcolsep}{1.2pt}\renewcommand{\arraystretch}{1.12}
\begin{tikzpicture}[remember picture]
\node[anchor=west,inner sep=0pt,outer sep=0pt] (feedback-grid) {\begin{tabular*}{\textwidth}{@{\extracolsep{\fill}}l*{6}{r}w{c}{7.5mm}w{c}{8mm}@{\hspace{4pt}}*{6}{r}w{c}{7.5mm}w{c}{8mm}@{}}
\toprule
& \multicolumn{8}{c}{\textbf{In-distribution (D1--D16)}} & \multicolumn{8}{c}{\textbf{Held-out (D17--D19)}}\\
\cmidrule(lr){2-9}\cmidrule(l){10-17}
\textbf{Method} & 160 & 320 & 480 & 640 & 800 & 960 & \makebox[7.5mm][c]{\tikz[remember picture,overlay]\coordinate (feedback-avg-id);\makebox[0pt][c]{\textbf{Avg.}}} & \makebox[8mm][c]{\tikz[remember picture,overlay]\coordinate (feedback-rank-id);\makebox[0pt][c]{\shortstack{\textbf{Avg.}\\\textbf{rank}}}} & 160 & 320 & 480 & 640 & 800 & 960 & \makebox[7.5mm][c]{\tikz[remember picture,overlay]\coordinate (feedback-avg-ood);\makebox[0pt][c]{\textbf{Avg.}}} & \makebox[8mm][c]{\tikz[remember picture,overlay]\coordinate (feedback-rank-ood);\makebox[0pt][c]{\shortstack{\textbf{Avg.}\\\textbf{rank}}}}\\
\midrule
CsiNet & \textcolor{blue}{\underline{-7.18}} & -9.44 & -10.39 & -11.05 & -11.33 & -11.65 & -10.17 & 3.11 & \textcolor{blue}{\underline{-5.75}} & \textcolor{blue}{\underline{-7.84}} & -9.00 & -9.50 & -9.92 & -10.33 & -8.72 & 3.00\\
TransNet & -7.11 & \textcolor{blue}{\underline{-9.99}} & \textcolor{blue}{\underline{-11.98}} & \textcolor{blue}{\underline{-13.09}} & \textcolor{blue}{\underline{-13.54}} & \textcolor{blue}{\underline{-13.99}} & \textcolor{blue}{\underline{-11.62}} & 2.66 & -5.74 & -7.76 & \textcolor{blue}{\underline{-9.12}} & \textcolor{blue}{\underline{-10.58}} & \textcolor{blue}{\underline{-11.42}} & \textcolor{blue}{\underline{-11.73}} & \textcolor{blue}{\underline{-9.39}} & 2.72\\
WiFo-CF  & -7.93 & -11.32 & -- & -13.70 & -- & -- & -- & \textcolor{blue}{\underline{2.04}} & -6.36 & -8.93 & -- & -11.23 & -- & -- & -- & \textcolor{blue}{\underline{2.11}}\\
WiFo-R & \textcolor{red}{\textbf{-8.58}} & \textcolor{red}{\textbf{-11.92}} & \textcolor{red}{\textbf{-13.88}} & \textcolor{red}{\textbf{-15.18}} & \textcolor{red}{\textbf{-15.69}} & \textcolor{red}{\textbf{-16.00}} & \textcolor{red}{\textbf{-13.54}} & \textcolor{red}{\textbf{1.21}} & \textcolor{red}{\textbf{-6.50}} & \textcolor{red}{\textbf{-9.09}} & \textcolor{red}{\textbf{-10.84}} & \textcolor{red}{\textbf{-12.03}} & \textcolor{red}{\textbf{-12.58}} & \textcolor{red}{\textbf{-12.99}} & \textcolor{red}{\textbf{-10.67}} & \textcolor{red}{\textbf{1.22}}\\
\bottomrule
\end{tabular*}};
\begin{scope}[overlay,on background layer]
\coordinate (feedback-band-top) at ([yshift=\dimexpr\ht\strutbox+\baselineskip\relax]feedback-avg-id);
\fill[black!10] ([xshift=\dimexpr-3.75mm-\tabcolsep\relax]feedback-avg-id |- feedback-band-top) rectangle ([xshift=\dimexpr3.75mm+\tabcolsep\relax]feedback-avg-id |- feedback-grid.south);
\fill[black!10] ([xshift=\dimexpr-3.75mm-\tabcolsep\relax]feedback-avg-ood |- feedback-band-top) rectangle ([xshift=\dimexpr3.75mm+\tabcolsep\relax]feedback-avg-ood |- feedback-grid.south);
\fill[red!8] ([xshift=\dimexpr-4mm-\tabcolsep\relax]feedback-rank-id |- feedback-band-top) rectangle ([xshift=\dimexpr4mm+\tabcolsep\relax]feedback-rank-id |- feedback-grid.south);
\fill[red!8] ([xshift=\dimexpr-4mm-\tabcolsep\relax]feedback-rank-ood |- feedback-band-top) rectangle ([xshift=\dimexpr4mm+\tabcolsep\relax]feedback-rank-ood |- feedback-grid.south);
\end{scope}
\end{tikzpicture}
\par\smallskip

\end{table*}

WiFo-R achieves average NMSEs of $-13.54$ dB on ID channels and $-10.67$ dB on OOD channels across the six payloads, outperforming the strongest task-specific baseline, TransNet, by $1.92$ and $1.28$ dB, respectively. As the feedback budget increases from 160 to 960 bits, the ID NMSE improves from $-8.58$ to $-16.00$ dB and the OOD NMSE improves from $-6.50$ to $-12.99$ dB. The progressive channel features therefore support increasingly accurate reconstruction across feedback rates, including unseen propagation configurations.

Importantly, all six feedback budgets reuse the same recurrent encoder, compression and reconstruction modules, and feedback decoder. A conventional per-scenario, per-rate deployment spanning 16 channel configurations and six payloads entails 96 model configurations, whereas WiFo-R jointly supports these settings with one rate-shared model. Compared with quantization-precision-based rate control, the proposed feature-length adaptation enables feedback-rate adjustment without changing the scalar quantization precision, substantially reducing the need for rate-specific model training and deployment.

\subsection{Transferability of Progressive Channel Representations}\label{subsec:transfer}
We evaluate the pretrained channel representations on two CSI-related inference tasks: wireless localization and LoS/NLoS classification. Localization is trained with 100 or 1000 labeled samples to test feature quality under different supervision budgets, while classification evaluates transfer across subcarrier configurations, including an unseen input resolution. WiFo-2 Base-Dense and WiT provide the localization references, and the classification results are compared with WiFo-2-Base-Dense and ResNet-3D.

\begin{figure}[t]
\centering
\includegraphics[width=\columnwidth]{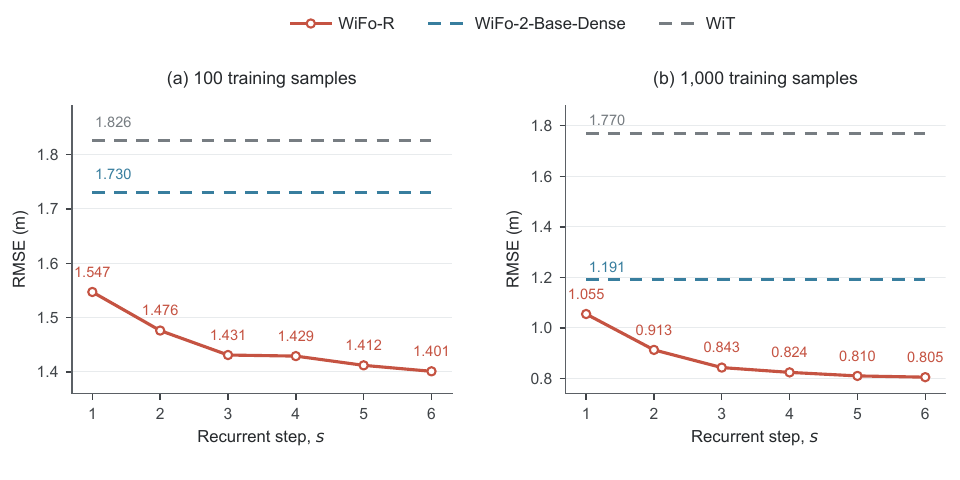}
\caption{Localization RMSE versus recurrent depth with (a) 100 and (b) 1,000 labeled samples.}
\label{fig:localization}
\end{figure}

\begin{figure*}[t]
\centering
\includegraphics[width=\textwidth]{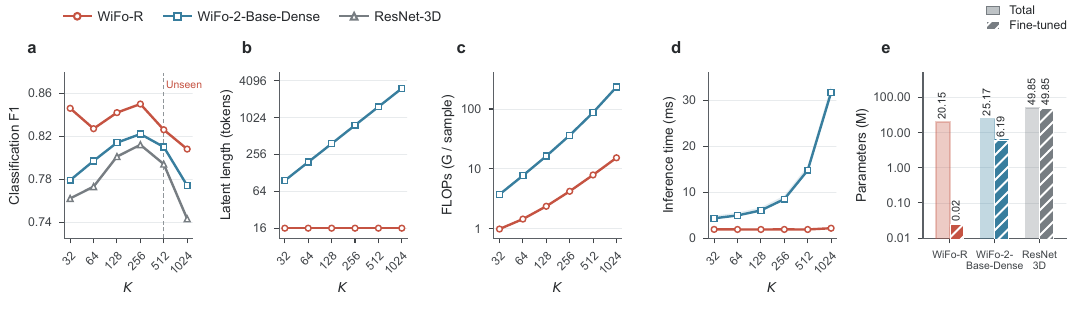}
\caption{LoS/NLoS classification and computational scaling: (a) macro-F1, (b) encoder output length, (c) FLOPs, and (d) inference latency versus subcarrier count; (e) total and fine-tuned parameters.}
\label{fig:classification-scaling}
\end{figure*}

Figure~\ref{fig:localization} shows a progressive reduction in localization RMSE as the recurrent depth increases from one to six steps: from $1.547$ to $1.401$ m with 100 training samples and from $1.191$ to $0.805$ m with 1000 samples. The compact one-step representation already outperforms both baselines, while deeper states further improve localization accuracy. This progression extends the benefits of recurrent refinement beyond CSI reconstruction, indicating that dominant location-related channel information is present at early depths and is enriched by subsequent updates. Figure~\ref{fig:classification-scaling}(a) further shows the strongest classification macro-F1 across the evaluated subcarrier counts, including 1024 subcarriers beyond the training range of at most 512. Together, these results demonstrate transferable channel representations that support progressive downstream localization and robust inference under changes in CSI observation resolution.

\subsection{Scalable Inference and Accuracy--Complexity Tradeoffs}\label{subsec:complexity}
We examine how WiFo-R scales with CSI observation resolution and recurrent depth. Figure~\ref{fig:classification-scaling}(b)--(d) compares latent sequence length, FLOPs, and batch-one latency for inputs of size $24\times8\times N_f$ as the subcarrier count increases. Measurements use FP32 inference on an NVIDIA RTX 4090 with 200 warm-up iterations and 1000 timed repetitions. At $s=1$, the encoder output remains at 16 channel tokens across the input range, while WiFo-2 Base-Dense reaches 4096 tokens at 1024 subcarriers. The cross-attention-based encoder consequently exhibits more favorable computational scaling and inference latency as the CSI observation size increases. Figure~\ref{fig:classification-scaling}(e) additionally reports the model-size and fine-tuning-parameter comparison for the classification task.

\begin{figure*}[t]
\centering
\includegraphics[width=\textwidth]{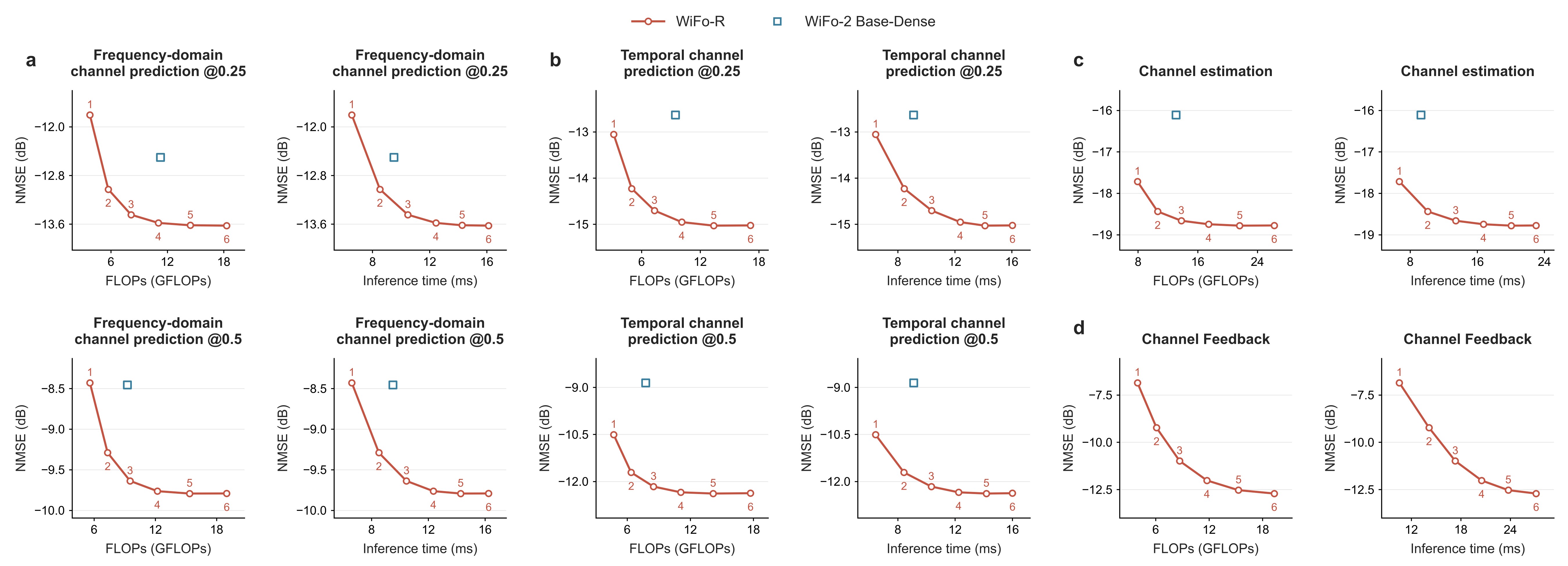}
\caption{Reconstruction NMSE versus FLOPs and inference latency across recurrent depths at 20-dB SNR: (a) CP-F, (b) CP-T, (c) CE, and (d) CF.}
\label{fig:complexity}
\end{figure*}

To quantify the cost of representation refinement, Fig.~\ref{fig:complexity} plots reconstruction NMSE against FLOPs and latency for all four reconstruction tasks. Each operating point corresponds to a different recurrent depth of the same trained model. The curves exhibit substantial early improvements followed by smaller gains at greater depths, with the preferred operating point depending on the CSI task. CP-T and CE reach strong accuracy at compact feature lengths, whereas CP-F and CF benefit from additional refinement when higher fidelity is required. This provides a direct mechanism for selecting the accuracy--computation trade-off without maintaining separate backbone models.

\subsection{Ablation Studies}\label{subsec:ablation}
% Single two-column IEEE floating table with two side-by-side panels.
% Transcribed from the author's supplied ablation screenshot. Lower NMSE is better.
\begin{table*}[!t]
\centering
\caption{Ablation study of architecture, pretraining, and task adaptation.}
\label{tab:ablation}
\vspace{2pt}
\begin{minipage}[t]{0.604\textwidth}
\centering
\fontsize{6.35}{7.75}\selectfont
\setlength{\tabcolsep}{2.0pt}\renewcommand{\arraystretch}{1.14}
\textbf{(a) CP-T and CP-F}\par\vspace{3pt}
\begin{tabular*}{\linewidth}{@{\extracolsep{\fill}}p{1.12in}ccccccc@{}}
\toprule
\multirow{2}{*}{\textbf{Variant}} & \multicolumn{2}{c}{\textbf{CP-T}} & \multicolumn{2}{c}{\textbf{CP-F}} & \multirow{2}{*}{\shortstack{\textbf{Params}\\\textbf{(M)}}} & \multirow{2}{*}{\shortstack{\textbf{FLOPs}\\\textbf{(G)}}} & \multirow{2}{*}{\shortstack{\textbf{Inference}\\\textbf{Time} \textbf{(ms)}}} \\
\cmidrule(lr){2-3}\cmidrule(lr){4-5}
& \textbf{ID} & \textbf{OOD} & \textbf{ID} & \textbf{OOD} & & & \\
\midrule
\textbf{Standard ($s=3$)} & \textcolor{blue}{\underline{-16.23}} & \textcolor{red}{\textbf{-18.03}} & \textcolor{blue}{\underline{-14.70}} & \textcolor{red}{\textbf{-16.02}} & \textcolor{red}{\textbf{26.40}} & \textcolor{red}{\textbf{9.31}} & \textcolor{red}{\textbf{12.99}} \\
Non-recurrent stacking & -15.06 & -15.81 & -14.41 & -15.11 & 87.85 & 11.21 & 14.61 \\
Self-attention variant & \textcolor{red}{\textbf{-16.52}} & \textcolor{blue}{\underline{-17.82}} & \textcolor{red}{\textbf{-15.04}} & \textcolor{blue}{\underline{-15.49}} & \textcolor{red}{\textbf{26.40}} & 15.32 & 18.95 \\
\addlinespace[1.5pt]
w/o dynamic-depth training & -12.26 & -14.93 & -10.26 & -12.51 & \textcolor{red}{\textbf{26.40}} & \textcolor{red}{\textbf{9.31}} & \textcolor{red}{\textbf{12.99}} \\
w/o random masking & -15.03 & -16.84 & -12.79 & -14.62 & \textcolor{red}{\textbf{26.40}} & \textcolor{red}{\textbf{9.31}} & \textcolor{red}{\textbf{12.99}} \\
\bottomrule
\end{tabular*}
\end{minipage}\hfill%
\begin{minipage}[t]{0.377\textwidth}
\centering
\fontsize{6.35}{7.75}\selectfont
\setlength{\tabcolsep}{2.0pt}\renewcommand{\arraystretch}{1.14}
\textbf{(b) CE and CF}\par\vspace{3pt}
\begin{tabular*}{\linewidth}{@{\extracolsep{\fill}}p{1.10in}cccc@{}}
\toprule
\multirow{2}{*}{\textbf{Variant}} & \multicolumn{2}{c}{\textbf{CE}} & \multicolumn{2}{c}{\textbf{CF}} \\
\cmidrule(lr){2-3}\cmidrule(l){4-5}
& \textbf{ID} & \textbf{OOD} & \textbf{ID} & \textbf{OOD} \\
\midrule
\textbf{Standard ($s=3$)} & \textcolor{red}{\textbf{-22.70}} & \textcolor{red}{\textbf{-20.87}} & \textcolor{red}{\textbf{-13.54}} & \textcolor{red}{\textbf{-10.67}} \\
Single-stage training & \textcolor{blue}{\underline{-21.94}} & \textcolor{blue}{\underline{-18.83}} & \textcolor{blue}{\underline{-12.07}} & \textcolor{blue}{\underline{-9.29}} \\
Scratch encoder & -20.14 & -17.45 & -10.01 & -8.57 \\
\addlinespace[1.5pt]
w/o LoRA & -21.59 & -18.52 & -11.38 & -9.12 \\
w/o token projector & -21.04 & -17.89 & -10.94 & -9.04 \\
\bottomrule
\end{tabular*}
\end{minipage}
\par\vspace{3pt}
\parbox{0.985\textwidth}{\raggedright\fontsize{7.0}{8.4}\selectfont
\emph{Note:} Model FLOPs and batch-one inference time are evaluated on D1 for CP-F with a 25\% masking ratio.\par}
\end{table*}

Table~\ref{tab:ablation} evaluates the recurrent encoder, cross-attention, multi-depth and multi-mask pretraining, and task-aligned adaptation. Compared with independently parameterized non-recurrent stacking, the shared recurrent model reduces the parameter count from 87.85M to 26.40M and improves OOD NMSE by $2.22$ dB for CP-T and $0.91$ dB for CP-F. Replacing cross-attention with self-attention increases FLOPs from 9.31 to 15.32 G and latency from 12.99 to 18.95 ms, while yielding lower OOD reconstruction accuracy. These comparisons confirm the efficiency and generalization advantages of the recurrent cross-attention encoder.

Multi-depth supervision is essential to progressive representation learning: removing it degrades ID CP-T and CP-F NMSE by $3.97$ and $4.44$ dB, respectively. Removing random masking further increases the errors by $1.20$ and $1.91$ dB, showing the complementary contributions of depth-wise supervision and heterogeneous masking objectives. Finally, the task-adaptation ablations establish the value of pretrained channel features. Compared with direct single-stage training, the full two-stage procedure improves OOD CE and CF NMSE by $2.04$ and $1.38$ dB, respectively; scratch initialization, removal of LoRA, and removal of feedback feature projection further degrade reconstruction quality.

\section{Conclusion}\label{sec:conclusion}
We introduced WiFo-R, a wireless foundation model with recurrent channel representation. Channel-query cross-attention separates CSI observation dimensions from the latent information interface, while parameter-shared recurrence progressively refines and expands that interface. Multi-mask, multi-depth pretraining produces useful channel features at several execution depths for prediction, estimation, feedback, and CSI-related inference. In variable-rate feedback, these recurrent states enable different quantized feature lengths using a unified model, with strong reconstruction accuracy and zero-shot transfer to OOD channel configurations. The task-dependent accuracy--capacity profiles and measured computational tradeoffs show that representation length can be selected according to the intended reconstruction quality and processing budget rather than fixed by the CSI sampling grid.

\bibliographystyle{IEEEtran}
\bibliography{WiFo-R}

@article{liu2026wifo2,
  author = {Boxun Liu and Xuanyu Liu and Shijian Gao and Xuesong Cai and Xiang Cheng and Liuqing Yang},
  title = {{WiFo-2}: A Generalist Foundation Model Unifies Heterogeneous Wireless System Design},
  journal = {National Science Review},
  volume = {13},
  number = {18},
  pages = {nwag500},
  year = {2026},
  doi = {10.1093/nsr/nwag500},
  url = {https://doi.org/10.1093/nsr/nwag500}
}

@article{liu2024wifo,
  author = {Boxun Liu and Shijian Gao and Xuanyu Liu and Xiang Cheng and Liuqing Yang},
  title = {{WiFo}: Wireless Foundation Model for Channel Prediction},
  journal = {Science China Information Sciences},
  volume = {68},
  number = {6},
  pages = {162302},
  year = {2025},
  doi = {10.1007/s11432-025-4349-0}
}

@article{alikhani2024lwm,
  author = {Sadjad Alikhani and Gouranga Charan and Ahmed Alkhateeb},
  title = {Large Wireless Model ({LWM}): A Foundation Model for Wireless Channels},
  journal = {arXiv preprint arXiv:2411.08872},
  year = {2024},
  url = {https://arxiv.org/abs/2411.08872}
}

@article{liu2026wifoinr,
  author = {Boxun Liu and Xuanyu Liu and Shijian Gao and Xiang Cheng and Liuqing Yang},
  title = {{WiFo-INR}: A Wireless Foundation Model Based on Implicit Neural Representations},
  journal = {arXiv preprint arXiv:2608.08033},
  year = {2026},
  url = {https://arxiv.org/abs/2608.08033}
}

@article{liu2025wifocf,
  author = {Xuanyu Liu and Shijian Gao and Boxun Liu and Xiang Cheng and Liuqing Yang},
  title = {{WiFo-CF}: Wireless Foundation Model for {CSI} Feedback},
  journal = {IEEE Transactions on Wireless Communications},
  volume = {25}, pages = {15039--15053},
  year = {2026},
  doi = {10.1109/TWC.2026.3681311}
}

@article{liu2025llm4wm,
  author = {Xuanyu Liu and Shijian Gao and Boxun Liu and Xiang Cheng and Liuqing Yang},
  title = {{LLM4WM}: Adapting {LLM} for Wireless Multi-Tasking},
  journal = {IEEE Transactions on Machine Learning in Communications and Networking},
  volume = {3}, pages = {835--847},
  year = {2025}, doi = {10.1109/TMLCN.2025.3585845}
}

@inproceedings{jaegle2021perceiver,
  author = {Andrew Jaegle and Felix Gimeno and Andy Brock and Oriol Vinyals and Andrew Zisserman and Joao Carreira},
  title = {Perceiver: General Perception with Iterative Attention},
  booktitle = {Proceedings of the 38th International Conference on Machine Learning},
  series = {Proceedings of Machine Learning Research},
  volume = {139},
  pages = {4651--4664},
  year = {2021},
  url = {https://proceedings.mlr.press/v139/jaegle21a.html}
}

@inproceedings{jaegle2022perceiverio,
  author = {Andrew Jaegle and Sebastian Borgeaud and Jean-Baptiste Alayrac and Carl Doersch and Catalin Ionescu and David Ding and Skanda Koppula and Daniel Zoran and Andrew Brock and Evan Shelhamer and Olivier Henaff and Matthew Botvinick and Andrew Zisserman and Oriol Vinyals and Joao Carreira},
  title = {Perceiver {IO}: A General Architecture for Structured Inputs \& Outputs},
  booktitle = {International Conference on Learning Representations},
  year = {2022},
  url = {https://arxiv.org/abs/2107.14795}
}

@article{graves2016act,
  author = {Alex Graves},
  title = {Adaptive Computation Time for Recurrent Neural Networks},
  journal = {arXiv preprint arXiv:1603.08983},
  year = {2016},
  url = {https://arxiv.org/abs/1603.08983}
}

@inproceedings{rao2021dynamicvit,
  author = {Yongming Rao and Wenliang Zhao and Benlin Liu and Jiwen Lu and Jie Zhou and Cho-Jui Hsieh},
  title = {{DynamicViT}: Efficient Vision Transformers with Dynamic Token Sparsification},
  booktitle = {Advances in Neural Information Processing Systems},
  volume = {34},
  year = {2021},
  url = {https://arxiv.org/abs/2106.02034}
}

@article{wen2018csinet,
  author = {Chao-Kai Wen and Wan-Ting Shih and Shi Jin},
  title = {Deep Learning for Massive {MIMO} {CSI} Feedback},
  journal = {IEEE Wireless Communications Letters},
  volume = {7}, number = {5}, pages = {748--751}, year = {2018},
  doi = {10.1109/LWC.2018.2818160}
}

@inproceedings{lu2020crnet,
  author = {Zhilin Lu and Jintao Wang and Jian Song},
  title = {Multi-resolution {CSI} Feedback with Deep Learning in Massive {MIMO} System},
  booktitle = {2020 IEEE International Conference on Communications (ICC)},
  pages = {1--6}, year = {2020},
  doi = {10.1109/ICC40277.2020.9149229}
}

@article{guo2020multirate,
  author = {Jiajia Guo and Chao-Kai Wen and Shi Jin and Geoffrey Ye Li},
  title = {Convolutional Neural Network-Based Multiple-Rate Compressive Sensing for Massive {MIMO} {CSI} Feedback: Design, Simulation, and Analysis},
  journal = {IEEE Transactions on Wireless Communications},
  volume = {19}, number = {4}, pages = {2827--2840}, year = {2020},
  doi = {10.1109/TWC.2020.2968430}
}

@article{lin2024scalable,
  author = {Yu-Chien Lin and Ta-Sung Lee and Zhi Ding},
  title = {A Scalable Deep Learning Framework for Dynamic {CSI} Feedback With Variable Antenna Port Numbers},
  journal = {IEEE Transactions on Wireless Communications},
  volume = {23}, number = {4}, pages = {3102--3116}, year = {2024},
  doi = {10.1109/TWC.2023.3305546}
}

@article{liu2024llm4cp,
  author = {Boxun Liu and Xuanyu Liu and Shijian Gao and Xiang Cheng and Liuqing Yang},
  title = {{LLM4CP}: Adapting Large Language Models for Channel Prediction},
  journal = {Journal of Communications and Information Networks},
  volume = {9}, number = {2}, pages = {113--125}, year = {2024},
  doi = {10.23919/JCIN.2024.10582829}
}

@inproceedings{he2022mae,
  author = {Kaiming He and Xinlei Chen and Saining Xie and Yanghao Li and Piotr Dollar and Ross Girshick},
  title = {Masked Autoencoders Are Scalable Vision Learners},
  booktitle = {Proceedings of the IEEE/CVF Conference on Computer Vision and Pattern Recognition},
  pages = {16000--16009}, year = {2022},
  doi = {10.1109/CVPR52688.2022.01553}
}

@inproceedings{sitzmann2020siren,
  author = {Vincent Sitzmann and Julien N. P. Martel and Alexander W. Bergman and David B. Lindell and Gordon Wetzstein},
  title = {Implicit Neural Representations with Periodic Activation Functions},
  booktitle = {Advances in Neural Information Processing Systems},
  volume = {33}, year = {2020},
  url = {https://papers.neurips.cc/paper_files/paper/2020/hash/53c04118df112c13a8c34b38343b9c10-Abstract.html}
}

@inproceedings{vaswani2017attention,
  author = {Ashish Vaswani and Noam Shazeer and Niki Parmar and Jakob Uszkoreit and Llion Jones and Aidan N. Gomez and Lukasz Kaiser and Illia Polosukhin},
  title = {Attention Is All You Need},
  booktitle = {Advances in Neural Information Processing Systems},
  volume = {30}, year = {2017},
  url = {https://papers.nips.cc/paper_files/paper/2017/hash/3f5ee243547dee91fbd053c1c4a845aa-Abstract.html}
}

@inproceedings{cai2020once,
  author = {Han Cai and Chuang Gan and Tianzhe Wang and Zhekai Zhang and Song Han},
  title = {Once for All: Train One Network and Specialize It for Efficient Deployment},
  booktitle = {International Conference on Learning Representations}, year = {2020},
  url = {https://openreview.net/forum?id=HylxE1HKwS}
}

@inproceedings{bolya2023tome,
  author = {Daniel Bolya and Cheng-Yang Fu and Xiaoliang Dai and Peizhao Zhang and Christoph Feichtenhofer and Judy Hoffman},
  title = {Token Merging: Your {ViT} but Faster},
  booktitle = {International Conference on Learning Representations}, year = {2023},
  url = {https://mlanthology.org/iclr/2023/bolya2023iclr-token/}
}

@inproceedings{ryoo2021tokenlearner,
  author = {Michael S. Ryoo and AJ Piergiovanni and Anurag Arnab and Mostafa Dehghani and Anelia Angelova},
  title = {{TokenLearner}: Adaptive Space-Time Tokenization for Videos},
  booktitle = {Advances in Neural Information Processing Systems}, volume = {34}, year = {2021},
  url = {https://papers.nips.cc/paper_files/paper/2021/hash/6a30e32e56fce5cf381895dfe6ca7b6f-Abstract.html}
}

@article{luan2023channelformer,
  author = {Dianxin Luan and John Thompson},
  title = {{Channelformer}: Attention Based Neural Solution for Wireless Channel Estimation and Effective Online Training},
  journal = {IEEE Transactions on Wireless Communications}, volume = {22}, number = {10},
  pages = {6562--6577}, year = {2023}, doi = {10.1109/TWC.2023.3244484}
}

@article{jiang2022transformer,
  author = {Hao Jiang and Mingyao Cui and Derrick Wing Kwan Ng and Linglong Dai},
  title = {Accurate Channel Prediction Based on Transformer: Making Mobility Negligible},
  journal = {IEEE Journal on Selected Areas in Communications}, volume = {40}, number = {9},
  pages = {2717--2732}, year = {2022}, doi = {10.1109/JSAC.2022.3191334}
}

@article{soltani2019channelnet,
  author = {Mehran Soltani and Vahid Pourahmadi and Ali Mirzaei and Hamid Sheikhzadeh},
  title = {Deep Learning-Based Channel Estimation},
  journal = {IEEE Communications Letters}, volume = {23}, number = {4},
  pages = {652--655}, year = {2019}, doi = {10.1109/LCOMM.2019.2898944}
}

@article{ye2018deepce,
  author = {Hao Ye and Geoffrey Ye Li and Biing-Hwang Juang},
  title = {Power of Deep Learning for Channel Estimation and Signal Detection in {OFDM} Systems},
  journal = {IEEE Wireless Communications Letters}, volume = {7}, number = {1},
  pages = {114--117}, year = {2018}, doi = {10.1109/LWC.2017.2757490}
}

@inproceedings{duggal2025alit,
  author = {Shivam Duggal and Phillip Isola and Antonio Torralba and William T. Freeman},
  title = {Adaptive Length Image Tokenization via Recurrent Allocation},
  booktitle = {International Conference on Learning Representations},
  year = {2025},
  url = {https://proceedings.iclr.cc/paper_files/paper/2025/hash/534ec8dfc0312d96e1c1b357304fadc2-Abstract-Conference.html}
}

@article{wang2026consisformer,
  author = {Yuwei Wang and Li Sun and Tingting Yang and Liwen Jing and Yuxuan Shi and Maged Elkashlan and Merouane Debbah},
  title = {{ConsisFormer}: Compute-Efficient Transformer for Wireless Foundation Models Based on Channel Consistency},
  journal = {arXiv preprint arXiv:2606.19953},
  year = {2026},
  url = {https://arxiv.org/abs/2606.19953}
}

@article{cui2022transnet,
  author = {Yaodong Cui and Aihuang Guo},
  title = {{TransNet}: Full Attention Network for {CSI} Feedback in {FDD} Massive {MIMO} System},
  journal = {IEEE Wireless Communications Letters},
  volume = {11}, number = {5}, pages = {903--907}, year = {2022},
  doi = {10.1109/LWC.2022.3149416}
}

@inproceedings{salihu2022wit,
  author = {Artan Salihu and Stefan Schwarz and Markus Rupp},
  title = {Attention Aided {CSI} Wireless Localization},
  booktitle = {2022 IEEE 23rd International Workshop on Signal Processing Advances in Wireless Communication (SPAWC)},
  pages = {1--5}, year = {2022},
  doi = {10.1109/SPAWC51304.2022.9833994}
}

@article{cheng2026fm4som,
  author = {Xiang Cheng and Boxun Liu and Xuanyu Liu and Ensong Liu and Ziwei Huang},
  title = {Foundation Model Empowered Synesthesia of Machines ({SoM}): {AI}-Native Intelligent Multi-Modal Sensing-Communication Integration},
  journal = {IEEE Transactions on Network Science and Engineering},
  volume = {13}, pages = {762--782}, year = {2026},
  doi = {10.1109/TNSE.2025.3587238}
}

@article{liu2026wifomisac,
  author = {Xuanyu Liu and Shijian Gao and Boxun Liu and Xiang Cheng and Liuqing Yang},
  title = {{WiFo-MiSAC}: A Wireless Foundation Model for Multimodal Sensing and Communication Integration via Synesthesia of Machines ({SoM})},
  journal = {arXiv preprint arXiv:2604.18255}, year = {2026},
  url = {https://arxiv.org/abs/2604.18255}
}

@article{wen2026wifoe,
  author = {Weibo Wen and Shijian Gao and Haotian Zhang and Xiang Cheng and Liuqing Yang},
  title = {{WiFo-E}: A Scalable Wireless Foundation Model for End-to-End {FDD} Precoding in Communication Networks},
  journal = {arXiv preprint arXiv:2601.09186}, year = {2026},
  url = {https://arxiv.org/abs/2601.09186}
}

@article{cheng2024som,
  author = {Xiang Cheng and Hong Zhang and Jun Zhang and Shijian Gao and Shuang Li and Ziwei Huang and Lixin Bai and Zheng Yang and Xin Zheng and Liuqing Yang},
  title = {Intelligent Multi-Modal Sensing-Communication Integration: Synesthesia of Machines},
  journal = {IEEE Communications Surveys \& Tutorials},
  volume = {26}, number = {1}, pages = {258--301}, year = {2024}
}

@article{cheng2025synthsom,
  author = {Xiang Cheng and Ziwei Huang and Yuming Yu and Lixin Bai and Ming Sun and Boxun Liu and Shuang Li and Jun Zhang and M. Lee},
  title = {{SynthSoM}: A Synthetic Intelligent Multi-Modal Sensing-Communication Dataset for Synesthesia of Machines ({SoM})},
  journal = {Scientific Data}, volume = {12}, year = {2025},
  note = {Article 819}
}

@article{zheng2026muse,
  title={{MUSE-FM: Multi-task environment-aware foundation model for wireless communications}},
  author={Zheng, Tianyue and Guo, Jiajia and Dai, Linglong and Jin, Shi and Zhang, Jun},
  journal={IEEE Transactions on Wireless Communications},
  year={2026},
  publisher={IEEE}
}
\end{document}